\documentclass[acmsmall]{acmart}
\acmJournal{PACMHCI}

\usepackage{xcolor}
\usepackage[utf8]{inputenc}
\usepackage{newunicodechar}
\newunicodechar{：}{:}
\usepackage{flushend}
\usepackage{float}
\usepackage{subfigure}
\AtBeginDocument{%
  \providecommand\BibTeX{{%
    \normalfont B\kern-0.5em{\scshape i\kern-0.25em b}\kern-0.8em\TeX}}}
\usepackage{tabularx}
\usepackage{bbding}
\usepackage{graphicx}
\usepackage{caption}
\usepackage{geometry}
\usepackage{amsmath}
\usepackage{makecell}
\usepackage{float}
\usepackage{color}
\usepackage{booktabs}
\usepackage[normalem]{ulem}
\usepackage{tabu}
\usepackage{hyperref}
\usepackage{xcolor}
\usepackage{appendix}
\usepackage{makecell}

\setcopyright{acmlicensed}
\acmJournal{PACMHCI}
\acmYear{2026} \acmVolume{10} 
\acmNumber{7} \acmArticle{GAMES002}
\acmMonth{11}\acmDOI{XXXXXXXXXXXXXXX}

\acmConference[CHI PLAY 2026]{CHI PLAY 2026}{November 2026}{York, UK}
\acmISBN{978-1-4503-XXXX-X/18/06}

\newif\iffinalversion
\finalversiontrue          
\iffinalversion
  \newcommand{\added}[1]{#1}       
  \newcommand{\deleted}[1]{}       
\else
  \newcommand{\added}[1]{\textcolor{blue}{#1}}
  \newcommand{\deleted}[1]{\textcolor{red}{\sout{#1}}}
\fi

\author{Shiyu Lei}
\authornote{Equal contribution.}
\orcid{0009-0005-7706-2028}
\affiliation{
\institution{City University of Hong Kong Studio for Narrative Spaces}
\city{Hong Kong}
\country{China}
}
\email{Leishiyu_@outlook.com}

\author{Ke-Xin Ren}
\authornotemark[1]
\orcid{0009-0008-1006-7125}
\affiliation{
\institution{City University of Hong Kong Studio for Narrative Spaces}
\city{Hong Kong}
\country{China}
}
\email{kexin.ren.ixd@gmail.com}

\author{ Daiyi Jiang}
\authornotemark[1]
\orcid{0009-0001-2651-0127}
\affiliation{
\institution{Department of Communication, University of Connecticut}
\city{Connecticut}
\country{United States}
}
\email{daiyi.jiang@uconn.edu}

\author{RAY LC}
\authornote{Correspondences can be addressed to ray.lc@cityu.edu.hk.}
\email{ray.lc@cityu.edu.hk}
\orcid{0000-0001-7310-8790}
\affiliation{
\institution{City University of Hong Kong Studio for Narrative Spaces}
\city{Hong Kong, SAR}
\country{China}}

\begin{document}

\title[Dragon Slayer Becomes the Dragon]{"Dragon Slayer Becomes the Dragon": How Players Perceive and Respond to Inequality in the Game World of Whiteout Survival}


\begin{abstract}
Inequality in real-world societies are associated with psychological distress and behavioral consequences. However, less is known about whether similar dynamics emerge when inequality exists within virtual environments or make-belief worlds. As online games increasingly constitute meaningful social spaces, it becomes critical to examine how players perceive and react to structural and resource differences online to optimize their experiences. This study studies perceptions of inequality in the online simulation game "Whiteout Survival," using semi-structured interviews and think-aloud gameplay walkthrough protocols. By focusing on players' interpretations of resource distribution, ranking systems, gaming mechanisms, and in-game social dynamics, our analyses revealed that players' attitudes on inequality vary according to their relative status: those occupying lower positions often criticize unfair structures, yet as they acquire stakes through resource accumulation or social integration, many defend the same systems they previously opposed. These shifts reveal how hierarchies reproduce position-dependent evaluations of fairness. The consequences of inequality on player actions depended on the transparency of game mechanisms, the structure of community hierarchies, and differential social capital. This work shows how human social perception and consequent actions are transformed when enacted in virtual processes in make-belief.
\end{abstract}




\maketitle

\section{Introduction}\label{sec:Introduction}
Inequality in real-world societies is widely associated with psychological strain and shifts in social cognition, including diminished sense of control, weakened institutional trust, and heightened sensitivity to unfair procedures \cite{ross2012sense,jordahl2009inequality,ariely2017corruption}. perceived inequality influences how individuals interpret fairness, environmental controllability, and future reward pathways \cite{amemiya2023thinking,knell2020perceptions}, often shaping risk-related decision-making and social judgment \cite{payne2017economic,pan2024does}. Yet structured inequality is increasingly embedded within virtual environments, particularly in large-scale multiplayer games where power, resources, and status are unevenly distributed by design. As games become a significant part of lived social experience, an important question emerges: when inequality is explicitly game-based, do these psychological and behavioral dynamics persist, or does the virtual context reshape how inequality is perceived and acted upon?

Inequality in games is related to, but not equivalent to inequality in real-world social systems. Prior work in virtual game contexts has examined how monetization shapes spending decisions, fairness judgments, and social relationships among players\cite{freeman2022pay}\cite{petrovskaya2022predatory}\cite{hamari2017}\cite{zhang2025microtransactions}, but less is known about how inequality operates in persistent game communities, where it becomes structured, long-term, and entangled with alliance membership, status, and social relationships \cite{freeman2022pay}\cite{petrovskaya2022predatory}. Existing work on in-game spending also does not fully explain how spending becomes connected to alliance power, responsibility, social status, and inequality \cite{hamari2017}\cite{zhang_looking_2015}. Similarly, while HCI research has studied social capital as a positive outcome of multiplayer game design and shown that toxic communities can weaken it, less is known about how players use social capital to respond to disadvantage and adapt to unequal systems \cite{frommel2023how}\cite{depping2018designing}.

To address these gaps and examine how inequality is perceived and enacted in game context, We selected Whiteout Survival as our research setting. As a globally popular mobile strategy game centered on alliance-based cooperation and pay-to-win progression, inequality in Whiteout Survival is multidimensional, cumulative, and socially institutionalized, echoing broader accounts of how status and hierarchy become embedded within structured social systems. This makes Whiteout Survival a useful site for examining how monetized inequality is perceived, negotiated, and sustained within a persistent game community.

To understand how players perceive and respond to inequality in this setting, we ask the following research questions (RQs):

\textbf{RQ1}: What factors contribute to the perception of inequality in a game community?

\textbf{RQ2}: What reported behaviors are engendered due to perceived inequality in this community?

\textbf{RQ3}: How is inequality sustained in this community? 

To answer these questions,we conducted remote semi-structured interviews and think-aloud game-play walkthrough with 11 players. Our analysis identifies two factors that shape players’ perception and evaluation of inequality (RQ1), four reported behavioral responses to perceived inequality (RQ2), and in-game community level processes through which inequality sustains itself, and parallel participation pathways that Sustain player's willing to stay Under Inequality (RQ3).

Based on our empirical analyses, we revealed that: (1) inequality perception in pay-to-win games is position-dependent; (2) the legitimacy of game inequality depends less on whether outcomes are equal and more on whether inequality is legible; (3) social capital functions as adaptive infrastructure under extreme game inequality; and (4) monetized inequality is maintained not only through game design, but also through community reproduction.

\section{Related Work}\label{sec:Background}
\subsection{Inequality and Its Psychological Effects}

Inequality is not only an objectively existing social phenomenon \cite{lee1990origin}, such as income disparities or class divisions, but also a factor that can be subjectively perceived by individuals \cite{kuhn2019subversive} and continuously trigger psychological reactions \cite{wilkinson2017enemy}. when individuals recognize their relatively disadvantaged position in the distribution of resources, status, or opportunities, this inequality transforms into psychological pressure through comparisons with their peers \cite{harth2008advantaged}. Inequality situations often accompany higher levels of anxiety and stress experiences \cite{charlton1997inequity}, making individuals more prone to feelings of deprivation, insecurity, and perceived threats from the environment \cite{burningham2003experiencing}. This psychological response manifests not only at the emotional level but also through stronger defensive attitudes and hostile tendencies \cite{foy2014emotions}, alongside diminished trust in others and the social environment \cite{jordahl2009inequality}. In other words, through individuals' comparative psychology \cite{ikeda2025perceived}, inequality causes them to persistently focus on their relative position, creating a latent state of psychological stress \cite{ridgeway2014status, achdut2023inequality}.

Perceived inequality not only triggers emotional responses but also shapes individuals' cognition and attitudes toward themselves, others, and their environment \cite{carlisle2008status, jetten2017social, tanjitpiyanond2022economic}. When individuals perceive themselves as being at a relative disadvantage through comparison, they tend to downgrade their assessment of their own capabilities, value, and development potential judgment of environmental controllability \cite{taylor1984theoretical, kraus2014undervalued, ross2012sense}. This transformation in limitations not stems from individual ability but from their understanding of resource distribution and social position \cite{abbott2006social}. In this process, inequality becomes internalized as a constraining background condition, leading individuals to attribute adverse outcomes to structural factors rather than their own capabilities \cite{amemiya2023thinking}.

Furthermore, perceived inequality also influences how individuals perceive others in their environment \cite{knell2020perceptions}. In inequality situations, people are more likely to interpret others' behaviors and intentions competitively or defensively \cite{sommet2019income}, viewing peers as potential competitors for resources or unfair beneficiaries, thereby reducing their tendency to make benevolent inferences about others' motivations \cite{ridgeway2014status}. This cognitive bias does not necessarily manifest as overt hostility \cite{glick2001ambivalent} but more likely appears as skepticism toward others' reliability and cooperative intent \cite{knell2021inequality}.

In individuals' understanding of environmental rules and social institutions, when they perceive resources or status as persistent and difficult to alter, they are more likely to perceive the rules themselves as biased or opaque \cite{goodwin1998situational}. This weakens their belief in the fairness of institutions and the effectiveness of norms \cite{ariely2017corruption}. Social rules may no longer be viewed as neutral mechanisms \cite{hurrell2005power} but rather understood as structural designs that perpetuate existing disparities \cite{patel2023addressing}.

In game scenarios, players do not automatically reject the inequality created by in-game monetization mechanisms, rather, they continuously assess whether these inequalities are reasonable, acceptable, and consistent with their expectations of fairness. Freeman et al. \cite{freeman2022pay} note that players do not view all forms of in-game purchases as unfair, instead, they judge them based on whether the advantages gained from such purchases disrupt the core game balance, undermine skill-based competition, or reduce player autonomy, it exhibits a highly contextual nature. Additionally, social monetization mechanisms are often deeply embedded in players' social relationships and community engagement. Player spending serves not only as a means to gain an advantage in the game but also as a mechanism to maintain social status, strengthen group identity, and avoid exclusion. Consequently, monetization not only creates economic disparities but also becomes embedded in social structures, transforming economic differences into socially meaningful hierarchical distinctions \cite{zhang2025microtransactions}. Moreover, players' perception of fairness is influenced by the structure of the game system. In multiplayer online competitive games, fairness is a core design objective that directly impacts player satisfaction, engagement, and retention. The game system shapes players' perception of fairness through matchmaking mechanism, character distribution, and interaction rules, making fairness an experience shaped by system design rather than merely an individual player's subjective assessment \cite{fan2024cupid}.

\subsection{Behavioral Consequences of Inequality}

Beyond macro-structural and psychological dimensions, inequality also influences individual behavior \cite{bapuji2015individuals}. Behavior is typically understood as an adaptive response to one's social position and environment structure \cite{eidelson1997complex}. When individuals perceive disparities in the distribution of resources, status, or opportunities through social comparison, this perception reshapes their understanding of environmental norms and others' intentions \cite{condon2020inequality, spence2011understanding}. This reshaping affects their judgment and choice within that environment.

Relevant research has explained behavioral differences among individuals in various inequality scenarios through perspectives such as relative position perception, social comparison, and understanding of environmental fairness \cite{fatke2018inequality, spence2011understanding, hochleitner2023inequality}. Within this framework, behavior is regarded as the externalization of an individual's perception of their social position in specific contexts \cite{korous2018unpacking}, rather than a simple direct reaction to objective resource disparities. Compared to other types of social behavior, risk-related decisions directly reflect how individuals perceive their position and environmental uncertainty in terms of the consequences of unequal actions \cite{fehr2025status, aven2013funtowicz}.

In terms of behavioral adaptation, players behavior is not determined solely by individual preferences but rather emerges as a dynamic adaptive process shaped by the interplay of personal characteristics and game structure. By tracking the long-term behavior of the same group of players in League of Legends and Team-fight Tactics, Chen et al. found that even though the two games encourage vastly different winning strategies, players still exhibit relatively stable behavioral patterns across different environments while adjusting their specific behaviors according to the game rules \cite{chen2026change}. 

Furthermore, game mechanics and system design can influence players' cognitive and behavioral adaptation processes. It has been noted that when faced with game challenges, failures, and setbacks, players complete tasks through continuous trial and error, learning, and strategy adjustment. Even though frustration is one of the most common negative emotions experienced during game-play, players are still able to gradually adapt to game rules through behavioral adjustments, thereby developing a higher level of engagement and a sustained willingness to participate \cite{sundaram2025frustration}. Similarly, when game content can be personalized based on player characteristics, players are more likely to experience satisfaction and engage on sustained participation. Compared to simple content adaption, the opportunity for players to make their own decisions and explore different development paths has a more significant impact on satisfaction \cite{auer2025enhancing}. 

Beyond the game mechanics, the game environment and interactive entities also shape players' behavioral adaption patterns. Players' perceptions of the transparency and fairness of punishment mechanisms influence their subsequent behavioral responses and coping strategies. When players perceive system rules as reasonable and procedurally fair, they are more likely to accept the rules and adjust their behavior \cite{ma2023transparency}. Meanwhile, Hochreiter et al. \cite{hochreiter2026beyond} found that NPC dialogue generated by LLM can enhance players' behavioral freedom and interaction flexibility, enabling them to exhibit more natural and proactive behavioral patterns when interacting with game characters. Similarly, in social simulation games, different perspective designs and social interaction structures influence players' cooperation, communication, and decision-making behaviors. Players adjust their strategies based on the information and social cues provided by the environment to achieve their goals more effectively \cite{sundaram2025frustration}. Similar observations have been reported outside gaming contexts. Zhang et al. \cite{zhang2025image} found that online dating users actively curate and reconstruct their profiles to fit platform norms and maximize desired outcomes. This suggests that adaptation to digital environments often involves strategic behavioral adjustment rather than passive compliance with system structures.

In terms of risk-related behaviors, when individuals perceive themselves as disadvantaged in social comparisons, it affects their cognition of future reward pathways \cite{pan2024does}, thereby influencing how they assess uncertain situations \cite{leuker2018exploiting}. From the perspective of environmental uncertainty and resource availability, unequal contexts reshape individuals' judgments about environmental stability and the predictability of rules \cite{kay2011social, murphy2018prejudiced}, transforming risk decisions into a response to structural uncertainty \cite{hoppner2026models}.

Furthermore, cognitive and social psychological research links risk-taking behaviors to factors such as perceptions of fairness \cite{mishra2015inequality}, sense of control \cite{damen2019sense}, and institutional trust \cite{dai2022paradoxical}. Risk-related behaviors do not stem solely from resource allocation or opportunity constraints, but rather from individuals' comprehensive judgments regarding the fairness of social rules, the reward potential of their efforts, and the reliability of social norms \cite{rabow1966role}. Consequently, risk-taking represents individuals' interpretations of the meaning within social structures, rather than direct reactions to objective disparities.

\subsection{Monetization, Fairness, and Predatory Design in Games}

Monetization is directly relevant to understanding inequality in games because it is the primary mechanism through which player hierarchies are created and maintained. When spending differences translate into structural advantages — shaping who holds power, who belongs, and who is excluded — players must negotiate fairness within a system that is designed to be unequal.

Prior research on monetization has developed along three interrelated  lines. First, Freeman et al.~\cite{freeman2022pay} showed that players of competitive online games make complex, context-dependent fairness judgments about in-game purchases, with acceptance depending on whether paid advantages disrupt core game-play balance. However, this work examines episodic, skill-based competitive games where fairness is evaluated at the level of individual matches --- it does not address how fairness judgments form and shift within persistent social hierarchies over months of play. Second, Petrovskaya and Zendle~\cite{petrovskaya2022predatory} categorized 35 monetization techniques players identify as unfair or aggressive, including engineered obsolescence, manufactured competitive pressure, and information asymmetry --- several of which appear explicitly in our participants' accounts. Third, Hamari et al.~\cite{hamari2017} identified six concrete motivational dimensions for in-game purchases in free-to-play games, finding that motivations related to \textit{unobstructed play} and \textit{social interaction} are significant predictors of how much players spend --- a finding particularly relevant to SLG (mobile simulation/ strategy game) contexts where alliance membership is structurally central and artificial barriers to progression are a primary monetization mechanism.

Beyond individual spending motivations, Zhang et al.~\cite{zhang2025microtransactions} extended this line of work by showing that predatory monetization in social games operates beyond individual financial harm: through interviews with 23 players of user-generated game platforms, they identified \textit{social monetization} which is a phenomenon in which in-game purchases become entangled with players' offline social relationships and community belonging. Their finding that players spend to maintain social standing and avoid exclusion rather than for game-play advantage establishes a conceptual bridge between monetization research and the social dynamics of hierarchical game communities. However, their study focuses on casual, platform-based games where social relationships are relatively transient; it does not examine how spending-driven social hierarchies become entrenched and self-reproducing within persistent alliance structures.

These studies on how game systems structure advantage, fairness, and player experience are important for understanding inequality perception in games. However, they largely examine monetization as a static phenomenon — focusing on individual purchase decisions or short-term fairness judgments in episodic game contexts. We should shift the perspectives of how inequality perception changes dynamically over time, as players move through persistent social hierarchies, accumulate stakes, and develop relationships within monetization-driven communities.

\subsection{Social Relationships and Hierarchies in Online and Gaming Communities}

With the expansion of the digital media environments, gaming communities have begun to be viewed by researchers as interactive spaces with social attributes \cite{ducheneaut2007virtual}. Online gaming environments are not only temporary, anonymous communications, but rather spaces capable of forming sustained interactive relationships, stable community structures over specific periods, and cooperative norms \cite{jia2015socializing, zakaria2022online, huang2025social}. Games can also function as narratives for social causes \cite{song_climate_2021,zhou_eternagram_2024,zhou_eternagram_2025, zhang_can_2025}. In World of Warcraft, users establish identity recognition, shared tasks, or long-term participation \cite{obst2018game, nardi2006strangers, debeauvais2011if}. These characteristics endow online games with the fundamental structural elements of real-world society.

Online gaming communities often maintain relatively stable membership structures and interaction networks \cite{ducheneaut2007life}. Relationships among players not only through real-time collaboration or competition but are also institutionalized and visualized through systems like friend networks, cooperative behaviors, and platform follow rules \cite{gong2024impact, bernal2025press, johansson2013if}. This relationship structure enables individuals to develop identifiable relative social positions within the community \cite{o2015sense}, such as core members, marginal participants, and opinion leaders, thereby forming a hierarchical community structure \cite{schiller2018inside}. Additionally, specific behavioral norms and cultural symbols exist within the community, such as “veterans” or “newbies” \cite{jin2017newbies}. These symbolic identities further reinforce the social attributes of online communities \cite{guegan2015online}. Therefore, online gaming communities function as digital social spaces with distinct social structures and interaction rules that may be modeled \cite{zhou_retrochat_2025}. Relationships, roles, and statuses among individuals are jointly shaped through continuous interaction and institutional design \cite{zervas2025digital, ducheneaut2004social}.

In online gaming environments, social relationships and hierarchical structures are constructed and manifested through distinct rules \cite{kolo2004living} and roles \cite{fu_i_2023}. For instance, in massively multiplayer online role-playing game (MMORPG), players collaborate in guilds, teams, or factions \cite{poor2014collaboration}, forming stable social relationships through teamwork and resource allocation \cite{chang2014team, nae2010dynamic}. Players possessing high-level gear or critical profession skills are more likely to occupy central positions in team decision-making or resource allocation, while newcomers or lower-level players tend to assume more dependent peripheral roles \cite{lisk2012leadership, ang2010social}.

In competitive online games, players’ status is typically evaluated by the system based on their win–loss records or skill levels and represented as visible ranks \cite{urbaniak2020identification}. This provides players with a clear relative position within the community, influencing their voice within teams and social standing \cite{jang2011exploring}. In games centered around social interaction, however, rank disparities stem more from social networks and resource control \cite{paavilainen2013social, lo2023ranking}. Some players become central nodes or opinion leaders within the community due to greater virtual assets, higher online activity, or stronger organizational abilities \cite{lee2013communication, manninen2007value, lisk2012leadership}. 

\subsection{Inequality Perception and Social Comparison}
Our analysis draws on two complementary theoretical frameworks to understand how players perceive and respond to inequality in gaming contexts.

\textbf{Payne's "Broken Ladder" Framework.} Payne (2017) argues that perceived inequality—individuals' subjective sense of where they stand relative to others—exerts more powerful psychological and behavioral effects than absolute poverty\cite{payne2017economic}. Using the metaphor of a "broken ladder," Payne describes how people constantly evaluate their position on social hierarchies through comparisons with reference groups. Critically, Payne demonstrates that only approximately 20\% of self-perceived status derives from objective markers like income or education; the remainder comes from subjective social comparison (p. 45). This perceived position triggers stress responses, shapes decision-making, and predicts behavioral adaptations including risk-taking, future discounting, and status competition. Payne's framework follows a process model: structural inequality creates visible disparities → individuals perceive their relative position → psychological stress/identity threats emerge → behavioral adaptations follow.

\textbf{Festinger's Social Comparison Theory.} Festinger (1954) proposed that humans possess an innate drive to evaluate their opinions and abilities (Hypothesis I), and when objective evaluation means are unavailable, they rely on social comparison (Hypothesis II). People preferentially compare themselves to similar others to achieve accurate self-evaluation (Hypothesis III), though there exists an upward drive to improve abilities that motivates comparison with superior others (Hypothesis IV). When differences become too large, comparison ceases because it no longer provides useful self-information (Hypothesis V)\cite{festinger1954theory}. Later extensions identified downward comparison comparing to those worse off, as a self-protective strategy when self-esteem is threatened.

\textbf{Integration for Gaming Contexts.} These frameworks together explain how inequality operates in digital games: Payne's model explains \textit{how inequality becomes visible} (through UI design, resource systems) and what behaviors result (risk-taking, withdrawal, competition), while Festinger's theory explains the psychological mechanisms mediating perception and behavior—specifically how players strategically choose comparison targets (upward/ downward/ horizontal) to manage self-evaluation and self-esteem. This integration allows us to trace the full process from structural game design through psychological processing to behavioral outcomes.

\subsection{Inequality Perception and Behavioral Response in Games}

In this work, we operationalize inequality as players’ perceived and experienced differences in resources, power, access, and social status that are structurally embedded in game systems and social organization, rather than momentary or incidental disparities. We focus on how such inequality is made visible through game mechanics and how players respond to it psychologically and behaviorally in everyday game-play.

Prior research has shown that inequality in real-world contexts shapes individuals' emotions, preferences, and decision-making behaviors, even when objective resource differences are small. However, less is known about how these processes unfold in persistent virtual environments. In multiplayer online games, forms of inequality commonly found in real-world societies are accelerated, amplified, and rendered highly visible through game systems, while also enabling responses that may not emerge under real-world constraints.

\section{Methods}\label{sec:Methods}
\subsection{Game Background Context}

Whiteout Survival is a mobile strategy game center on a glacial apocalypse survival theme released in February 2023 by Century Games \citep{whiteoutsurvival}. The game appeals to a wide global audience, by 2025, the total download volume on Google play and apple app store exceeded 90 million. In 2025, the game reached the top-grossing position globally and its total revenue exceeded 140 million dollars \citep{businessofapps_topgrossinggames}  \citep{activeplayer_whiteoutsurvival}. 

To examine the perception of inequality in game-play contexts and its influence on player decision-making, this study selects Whiteout Survival as the research setting. Compared with game genres such as MMORPGs or MOBAs, which center on individual mechanical skill, moment-to-moment decision-making, or stage-based progression, inequality in these games is often structured around players’ technical proficiency, time investment, or game knowledge, and is partially reset at the end of each match or season.
In contrast, as a mobile SLG (mobile simulation /strategy game), Whiteout Survival uses game mechanics, monetary investment, large-scale cooperation, social status, and long-term time accumulation into a continuous game-play process, resulting in multidimensional, long-term, and structural differentiation in player strength. 

\begin{figure}
    \centering
    \includegraphics[width=1\linewidth]{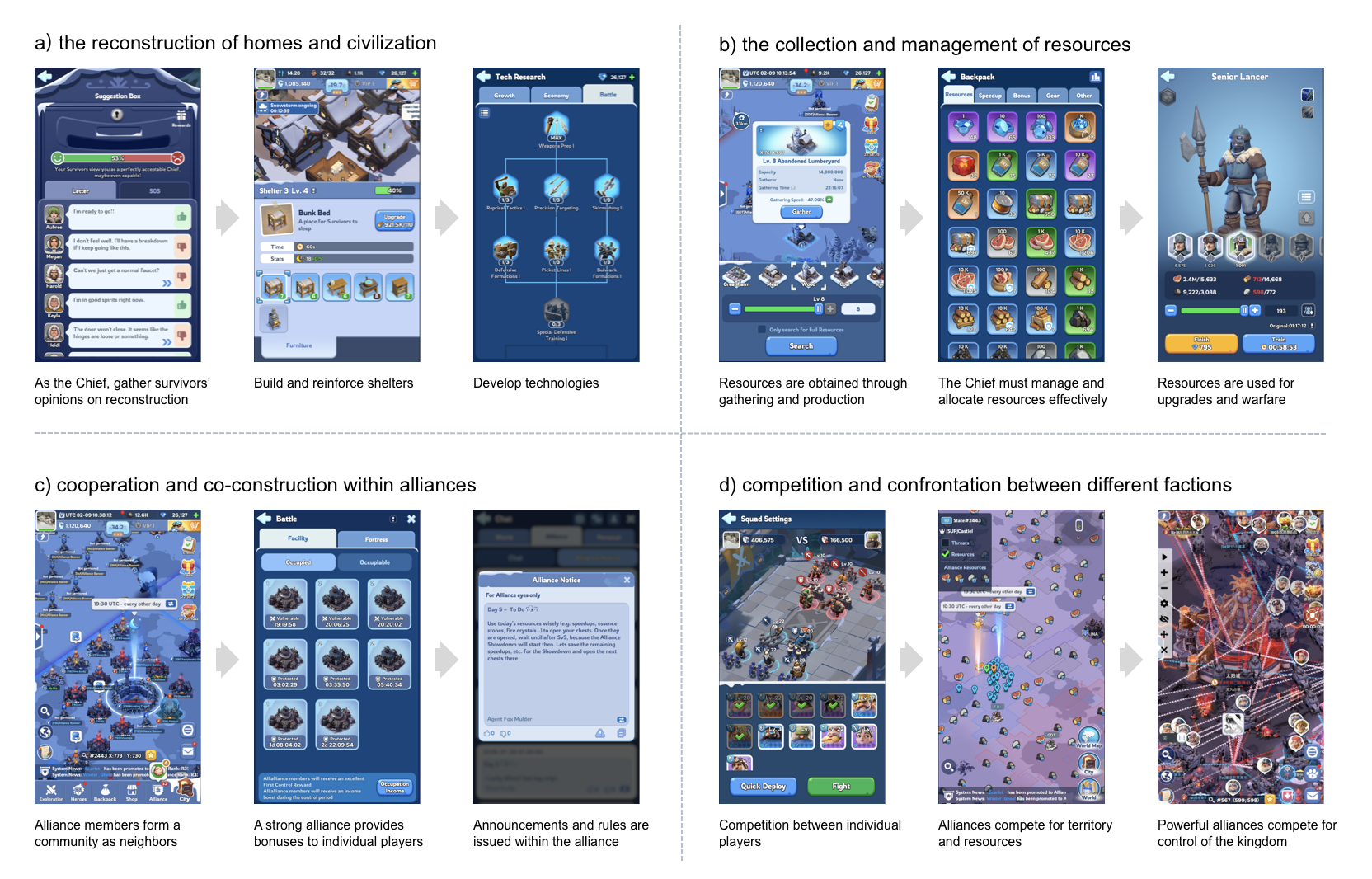}
    \caption{Whiteout Survival combines idle mechanics with strategic decision-making. The’ primary game activities of the players include: a) the reconstruction of homes and civilization within a post-apocalyptic survival setting; b) the collection and management of resources such as materials and military forces; c) cooperation and co-construction within socially driven alliances; and d) competition and confrontation between different factions, including individual players, alliances, and nations.}
    \label{fig:placeholder}
    \end{figure}

In this game, alliances serve as the primary social units through which players coordinate resource acquisition, collective combat, and participation in large-scale events. Alliances operate under a clearly defined hierarchical structure consisting of five ranks (R1–R5). Rank five (R5) represents the alliance leader, who holds overarching decision-making and administrative authority, while rank four (R4) members act as officers appointed to support alliance management and coordination. Ranks one to three (R1–R3) correspond to regular members with progressively fewer permissions. Leadership positions are not determined solely by combat power; instead, they often depend on players’ availability, sustained activity, and ability to coordinate others. In practice, alliances also develop a functional division of labor, in which high-power players primarily assume frontline combat roles, mid-power players focus on execution and support tasks, and lower-power players contribute mainly through construction and resource development.

Inequality is repeatedly and strongly perceived during the game, as most attributes, power levels, ranks, progression milestones, and match outcomes are made highly transparent through ranking systems. This high visibility intensifies players’ awareness of competition and disparity, thereby providing an appropriate context for examining how inequality is perceived and translated into concrete decision-making.

\begin{figure}
    \centering
    \includegraphics[width=0.5\linewidth]{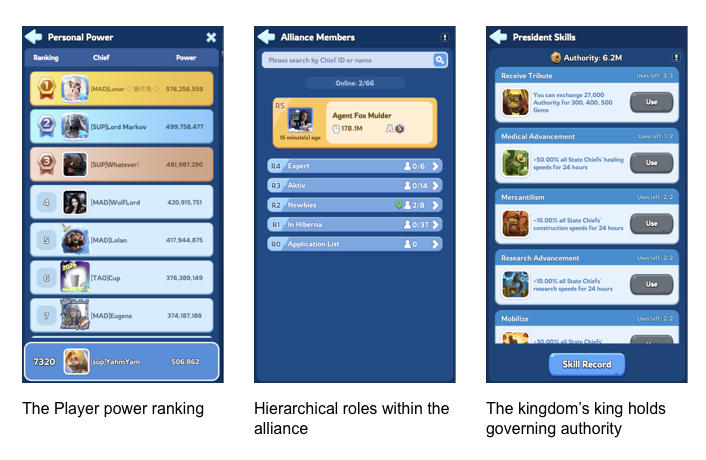}
    \caption{Forms of player inequality in Whiteout Survival}
    \label{fig:placeholder}
\end{figure}

Taking multiplayer online games such as "Whiteout Survival" as the research context, the inequality phenomenon therein not only reflects the unequal structure in the real society to a certain extent, but is also significantly magnified and accelerated through the design of the game mechanism. Compared with the situation in the real society where inequality often accumulates gradually over a long time scale and is difficult to be fully observed, games embed players' perception of inequality, emotional responses, and decision-making behaviors into a clearer and traceable action path, enabling researchers to observe how inequality is recognized and transformed into specific behaviors within a relatively compressed time frame. Meanwhile, compared with the real society, the consequences of risky behaviors in games are usually reversible and the risk cost is relatively low. This feature lowers the psychological threshold for players to test, resist or make strategic adjustments, thus making it easier for them to reveal their intuitive reactions and decision-making logic in unequal situations. Therefore, the game scenarios represented by Whiteout Survival provide a research environment that is difficult to obtain in reality, observable and comparable for studying the perception and decision-making mechanisms of inequality.

\subsection{Participants}

We recruited participants through an online questionnaire collecting demographic information, gaming backgrounds, and Whiteout Survival experiences. The questionnaire used a bilingual format (English and Mandarin Chinese). We deliberately avoided using the term "inequality" in questions to prevent priming effects, instead asking about game-play experiences, alliance dynamics, spending patterns, and social interactions using neutral language. The questionnaire took approximately 6 minutes to complete. We advertised the study through Whiteout Survival player communities on Discord, Reddit (r/whiteoutsurvival), Taptap, Rednote, and Wechat group. Snowball sampling was also used, as some participants referred additional players, who were then asked to complete the same questionnaire and were screened using the same inclusion criteria. We attempted to recruit a participant sample with diversity regarding gender, age, occupation and game experience, but we did not apply any further inclusion or exclusion criteria in the recruitment process, all eligible survey respondents were invited to participate in an interview.

Our participant inclusion criteria were: (1) active Whiteout Survival players with at least three months of game-play experience, (2) at game level 19 or above and (3) adults aged 18 years or older. Participants who completed both the online survey and remote interview received 5 dollars USD or equivalent in compensation. We received 26 responses, of which 25 met the inclusion criteria, one response was excluded because the player’s level was below 19. All 25 eligible respondents were invited to participate in follow-up interviews, 11 responded and were interviewed.

Out of the 11 interviewees (age 19 to 33; 63.6\% female, 36.4\% male), we included 2 free-to-play players, 4 light spenders, 2 moderate spenders, 2 heavy spenders, and 1 very high spender. Our sample captured diverse alliance positions: 5 managers/ leaders, 5 regular members, and 1 solo player. Play duration ranged from 3–6 months (5 participants), 7–12 months (1 participant), to over 1 year (5 participants), with weekly playtime varying from 3-6 hours (2 participants) to 21+ hours (3 participants). Most participants (9/11) had reached Great Hearth Level 31+, indicating experienced mid-to-late game players. Table 1 summarizes this variation across 11 participants. 
\begin{table*}[h]
\begin{tabular}{llllllll}

\makecell[l]{\textbf{ID}\\\textbf{ }}  & \makecell[l]{\textbf{Gender}\\\textbf{ }}& \makecell[l]{\textbf{Age}\\\textbf{ }}&\makecell[l]{\textbf{Level}\\\textbf{ }}&\makecell[l]{\textbf{Play}\\\textbf{Duration}}&\makecell[l]{\textbf{Total}\\\textbf{Spending}} &\makecell[l]{\textbf{Alliance}\\\textbf{Role}} &\makecell[l]{\textbf{Weekly}\\\textbf{Hours}}\\ 

P01& Female& 20& 31+ & 1 year+& 620& \added{R5} (main manager) &7-13 \\ 
P02& Male&  20& 31+ & 1 year+& 5& \added{R1} (regular player)&21+ \\ 
P03&  Female& 22& 31+ & 7-12 months& 1440& \added{R3} (regular player)&7-13 \\
P04&  Male& 30& 31+& 1 year+& 2890& \added{R4} (high-power player)&14-21 \\

 P05& Female& 30& 31+ & 1 year+& 720& \added{R4} (manager)&21+\\
 P06& Male& 21& 31+& 3-6 months& 290& solo player&3-6 \\
 P07& Female& 25& 31+& 1 year+& 145&  \added{R5} (main manager)&14-21 \\
 P08& Male& 33& 22-24& 3-6 months& 0& \added{R2} (regular player)&3-6 \\
 P09& Female& 23& 31+& 3-6 months& 1445& \added{R4} (manager)&7-13 \\
 P10& Female& 22& 31+& 3-6 months& 145& \added{R3} (regular player)&21+ \\
 P11& Female& 19& 28-30& 3-6 months& 200& \added{R4} (manager)&7-13 \\
 & & & & & & &\\
\end{tabular}
  \caption {Demographic Information of Participants. Level refers to Great Hearth Level, which is participants' in-game progression milestone. Total Spending refers to cumulative in-app purchases in USD. Alliance Role refers to participants' position within the alliance hierarchy, which consists of five ranks (R1--R5): managers hold leadership responsibilities, high-power players primarily contribute to combat, regular players are standard alliance members, and solo players do not have a stable alliance. P06 is listed as a solo player and therefore did not hold an alliance role.}
  \label{table1}
\end{table*}

\subsection{Interview and Game-play Walkthrough}

Three researchers first jointly conducted two interviews to align interviewing practices. After discussion and calibration, the three researchers then independently conducted the remaining nine interviews using the same interview protocol. Each interview session lasted approximately 65 minutes total, consisting of three parts: a 40-minute semi-structured interview, a 5 minute break, and a 20-minute think-aloud game-play walkthrough. All interviews were conducted remotely via video conferencing platforms (Zoom, Tencent Meeting), and voice-recorded with participants’ permission, all interviews were transcribed and then checked by the research team. The interviews were conducted in English or Chinese, with the transcripts translated into English (DeepL) if the interview was in Chinese.

\textbf{Semi-Structured Interview.} To obtain an in-depth understanding of participants’ perceptions of inequality and their behavioral responses, we asked semi-structured questions covering four main areas: (1) players’ overall game-play experiences and progression in Whiteout Survival, (2) their alliance dynamics and interactions with players at different resource levels, (3) specific experiences or memorable moments involving resource differences between players, and (4) any changes in their play patterns, strategies, or attitudes over time. We included the semi-structured interview protocol in Appendix A.

\textbf{Think-Aloud Game-play Walkthrough.} After the interview and 5 minute break, participants were asked to share their mobile screens via the video conferencing platform and provide a guided tour of their game-play experience. Using their own devices (personal smartphones), participants demonstrated their daily routines in Whiteout Survival, including checking alliance messages, reviewing ranking boards, examining their base development, and navigating social features. Participants were encouraged to show us specific game features or social interactions they had mentioned during the interview. While some participants occasionally engaged in brief combat scenarios or resource management tasks, the primary focus was on demonstrating social interfaces, communication channels, and progression indicators rather than active game-play. 

We asked participants to verbalize their thoughts while performing these activities. When they paused or became silent, we prompted them with questions such as “What are you looking at here?” and “What are you thinking about now?” When they interacted with inequality-relevant game elements, such as power rankings, alliance member lists, spending-related features, alliance structures, and chat windows across different channels, the researcher occasionally asked clarifying and follow-up questions about visible game elements. These questions included “How do you interpret these power differences?”, “What does this ranking mean to you?”, “Why do you check this feature?”, and “What does this information tell you?” These prompts helped us understand participants’ real-time sense-making of inequality-related information. The think-aloud game-play walkthrough triggered situated reflections through in-game cues and helped participants articulate inequality-related experiences that were not always recalled during the interview\citep{ericsson1993protocol}\citep{holtzblatt1993contextual}.

Screen recordings of the game-play walkthrough and audio from both the interview and walkthrough were captured with participants’ consent. Recordings were transcribed and translated into English. 

\subsection{Data Analysis}

Interview and think-aloud game-play walkthrough transcripts were analyzed together as the primary coding corpus. We used inductive thematic coding guided by Payne’s perceived inequality framework\cite{payne2017economic} and Festinger's Social Comparison Theory\cite{festinger1954theory}to identify patterns in how players perceived, interpreted, and responded to inequality in Whiteout Survival. They helped us attend to how players perceived inequality through structural visibility and social comparison, processed these perceptions through comparison target selection and identity negotiation, and adapted behaviorally through strategic responses.

In the early coding stage, two researchers conducted initial and focused coding iteratively. Initial coding helped us identify concepts that appeared in participants’ accounts, while focused coding allowed us to prioritize codes most relevant to players’ perceptions of inequality and behavioral responses. Before moving to higher-level categorization, the two researchers independently coded a shared subset of the corpus, consisting of the interview and think-aloud transcripts from R01 and R02, to assess agreement on the emerging coding scheme.
Cohen's Kappa was calculated at this stage ($\kappa$ = .76), indicating substantial agreement\cite{mcdonald2019reliability}\cite{Cohen1960ACO}\cite{landis1977measurement}. The researchers then discussed coding differences and refined the coding scheme before applying it to the remaining transcripts. In the axial coding stage, related codes were grouped into broader categories by examining relationships between them. In the selective coding stage, these categories were first synthesized into five overarching themes, then we revisited the coded data and refined the themes iteratively into six themes structuring the Results section, to better align with the revised research questions, following thematic analysis practices that treat theme development as an iterative process\cite{braun2006using}. Emerging codes, categories, and themes were reviewed and refined through discussions among all four members of the research team.

The interview questions were organized around four areas related to the inequality perception-response process: (1) mechanisms through which inequality becomes perceptible (visibility, information access, legitimacy assessment), (2) psychological responses mediating perception and behavior (emotional trajectories, identity negotiation, cognitive reframing), (3) behavioral adaptation patterns (intensification, attrition, social hedging, compensatory risk), and (4) cross-context response comparisons (mirroring vs. disconnect between virtual and real-world inequality responses).

\section{Result}\label{sec:Result}
We report six themes organized around three research questions. RQ1 examines how players perceive and evaluate inequality — shaped by their position in the hierarchy and the transparency of its mechanisms. RQ2 examines how players respond — mediated by alliance social dynamics and falling along a spectrum from strategic adaptation to compensatory aggression. RQ3 examines how inequality is sustained — through socialization that converts critics into defenders, and alternative pathways that give players genuine reasons to stay . Together these themes trace inequality from individual perception through collective behavior to community level reproduction.

\subsection{What factors contribute to the perception of inequality in a game community}

Players' perceptions of inequality were not determined primarily by the magnitude of resource or power gaps, but rather by two distinct factors: their relative position within the game's social hierarchy, and the transparency of the mechanisms producing those gaps. We first examine how status position shapes fairness evaluations (4.1.1), then analyze how transparency functions as a legitimacy mechanism (4.1.2).

\subsubsection{Theme1 Status Position Shapes Fairness Evaluation: "The Dragon Slayer Becomes the Dragon"}

Players' evaluations of inequality transformed as their positions within the game hierarchy changed. This shift occurred not through slow socialization but through rapid stake acquisition—a dynamic unique to digital environments where resource accumulation and social integration can happen within weeks rather than years.

P06 exemplified this transformation. When he first joined the server, he unknowingly violated an unwritten rule—attacking other players' mining operations. Stronger players immediately "zeroed" him (confiscated all his resources), leaving him unable to continue playing: "\textit{I didn't understand the rules then and accidentally broke some 'default rules' of the server. As a result, I was completely 'zeroed' by stronger players, that was when I felt how much the power gap affects the game experience. I was very angry.}" Six months later, having accumulated resources and integrated into the server community, P06's evaluation had completely reversed:"\textit{Now that I've played longer, I actually agree with these rules because if no one hits mines, no one loses troops or resources. It makes the game more peaceful.}" When asked about his current feelings toward the game's top players, he no longer expressed resentment: "\textit{I don't feel the rules are controlling me; they are there to make the game environment better.}" The rules had not changed. His position had.

Two mechanisms drove this transformation. First, \textbf{resource accumulation} created stakes worth protecting. As players gathered resources over time, the same rules that once punished their ignorance now protected their investments. P06's shift occurred after he had accumulated enough resources that the "no hitting mines" rule benefited rather than harmed him. His current game-play focused on protecting what he had built rather than rapid expansion, making stability-preserving rules attractive. Second, \textbf{social integration} provided a new interpretive framework. P01 described this shift: \textit{"If you don't participate in these social circles — like not joining the WeChat group — the game feels like a single-player experience. Once you join the social circle, the experience changes completely; you feel like you are fighting for a group of people rather than just for the game mechanics."} Joining the alliance's social network meant adopting the alliance's evaluative lens — rules were now judged by collective benefit rather than individual advantage. P06 eventually participated in enforcing the same rules that had once punished him, a pattern we return to in Theme 3.

What made this dynamic distinctive was its \textbf{speed of position transformation in a virtual game environment}. Unlike material hierarchies where status acquisition occurs over years or generations, digital environments enabled rapid transformation. P06's shift from angry victim to grateful defender occurred within six months. P07's group moved from oppressed to oppressors within weeks of server migration. This temporal compression occurred because digital resource accumulation happened faster (daily login rewards, event participation) and social integration required lower barriers (joining WeChat groups, participating in voice chat during alliance events). One could move from outsider to insider not through years of relationship-building but through weeks of active participation and visible contribution.

\textit{The "Dragon Slayer" Pattern: From Oppressed to Oppressor}; The most dramatic manifestation appeared when entire groups underwent status shifts. P07 described what she called "the dragon slayer becoming the dragon": "\textit{We originally moved because we were oppressed by 'Earth-tier' whales (extremely high power) in our old state and felt the distribution was unfair. But in the new state, when we became the strongest, the people who used to cry for fairness created even more restrictive rules to take almost all the rewards for themselves. I found this greed intolerable}." This pattern revealed that position change didn't just shift individual attitudes—it could transform entire groups from victims advocating fairness into elites imposing even harsher inequality. P07 immediately migrated again, demonstrating that not all players internalized new positions, but the pattern was widespread enough to constitute a recognizable dynamic.

Fairness perception is not a judgment players make from outside the system — it is a judgment they make from inside it, from wherever they currently stand. The same rules, the same gaps, the same hierarchy produce opposite verdicts depending on one's position. Position is not background context; it is the primary variable. 

\subsubsection{Theme2  Players Perceive That Transparency in the System Can Legitimize Inequality
}
Beyond status position, players evaluated inequality through whether they could understand and predict how advantages were obtained. Transparency—the visibility and intelligibility of progression mechanisms—emerged as the primary determinant of legitimacy. When cost structures were clear, even extreme inequality became understandable and tolerable. When mechanisms were opaque, inconsistent rule application , or manipulative, legitimacy collapsed regardless of gap size.

\textit{How Transparency Manifests in Player Behavior}; Players evaluated inequality not by measuring gaps but by assessing whether they could understand and predict how advantages were obtained. This manifested in specific information-seeking behaviors and spending decisions. P04 described his early game-play: "\textit{After playing for a long time, you understand that many things in this game are essentially 'how much money for what result.' Once you know how much something costs, you actually become less impulsive about it.}" Knowing that exclusive content required hundreds of thousands of RMB in sustained spending led not to competitive aspiration but to what P04 called "upward severing" — a deliberate psychological disconnection from tiers he recognized as unreachable. Transparency gave players a rational stopping point. Instead of sustained frustration, they disengaged from comparisons that no longer served them. This is precisely the dynamic Festinger described: when a gap becomes too large to provide useful self-information, comparison ceases. In Whiteout Survival, transparency was what allowed players to recognize when that threshold had been crossed.

When mechanisms were clear, players made calculated purchases. P03 described strategic early spending: "\textit{My first purchase was for an extra building queue. Early on, this game focuses heavily on city building, and having more queues saves a lot of time.}" Later, she spent strategically during events: "\textit{During combat-heavy stages, my combat power was relatively low, but by spending money and studying the mechanics, I could defeat players with higher power. For example, in the Burning Mine event, my power was only mid-tier, but I could rank in the top three. That kind of experience made me feel the investment was worthwhile}." Transparency transformed spending from emotional impulse into calculated agency.

While transparency legitimized extreme inequality, three conditions destroyed legitimacy regardless of gap size: opacity, inconsistent rule application, and deliberate manipulation. The first was \textbf{opacity through information gate-keeping}. The game's lack of in-game communication created hidden information channels. P01 identified this as the most significant structural problem: "\textit{From the start, there are 'veterans' who come from older servers and know the ropes; they proactively create WeChat groups and recruit people. If you don't participate in these social circles—like not joining the WeChat group—the game feels like a single-player experience.}" Critical coordination information—event strategies, diplomatic negotiations, resource management—circulated in private WeChat groups controlled by established players. When asked what single mechanism she would most change, P01 cited precisely this: "\textit{I most want to change the 'Information Gap.' I hope the developers could provide a more open, unified communication platform.}" Players didn't object to whales having more power; they objected to not knowing \textit{how} to optimize their own play because essential information was gate-keeping. 

The second condition was \textbf{inconsistent rule application}. As Theme 1 showed, players who gained power frequently imposed harsher rules than those they had previously condemned. What destroyed legitimacy here was not the inequality itself but the broken consistency — the implicit contract that rules apply predictably to everyone, regardless of position. When that contract was violated, players lost not just trust in specific individuals but confidence in the alliance and rule system as a whole.

Thirdly, \textbf{deliberate manipulation}. P09 identified developer-planted "Shills" (provocateurs): "\textit{They use 'Shills' to provoke Whales into fighting... They intentionally design these unequal mechanisms to maximize profit.}" Manufactured conflict—developers creating artificial competitive pressure through planted actors—violated the "price-for-value" contract. Players expected to pay for power, not to be psychologically manipulated into competitive spending through fake opponents. Similarly, P04 identified engineered obsolescence: "\textit{Each new generation of heroes has higher stats, so if you want to keep your advantage, you must keep spending every generation. The money you spent before keeps losing value, which creates a strong sense of loss.}" \textit{When asked if developers tried to reduce inequality: "Yes, but the goal is not fairness—it is continuous revenue}." Recognition of these mechanisms led not to continued competitive spending but to strategic disengagement.

Spending accumulation in Whiteout Survival happened gradually and often without players fully registering the total. Table 1 shows that several participants had spent over \$1,000 — P09 spent approximately \$1,445 over three to six months, roughly \$240–480 per month; P04 spent \$2,890 over more than a year; P03 spent \$1,440 over seven to twelve months. P03 described the psychological mechanism behind this accumulation: \textit{"Spending money itself is a kind of 'icebreaker' — once you make the first purchase, it's easier to make the second."} Players would reach 90\% of an event milestone and face an incremental spending decision that felt justified in the moment while accumulated totals became substantial. P04 identified a structural driver which is engineered obsolescence: \textit{"Each new generation of heroes has higher stats, so if you want to keep your advantage, you must keep spending every generation. The money you spent before keeps losing value."} Maintaining relative position, not gaining absolute power, required perpetual investment. For players who recognized this treadmill, the response was not escalation but deliberate disengagement as P04 described "upward severing," a conscious decision to stop comparing upward once the cost of top-tier status became fully visible.

These patterns reveal that players applied a coherent, if unstated, standard when evaluating inequality: legitimate inequality is legible inequality — visible, predictable, and consistently applied. Opacity, inconsistent enforcement, and deliberate manipulation each violated one of these conditions, and each was sufficient on its own to collapse legitimacy. This finding extends RQ1 beyond what Theme 1 established: position tells players where they stand in the hierarchy, but transparency tells them whether standing there makes sense. Together, these two factors form the perceptual foundation on which all behavioral responses are built — which is where we turn next.

\subsection{What reported behaviors are engendered due to perceived inequality in this community}

This part examines how players translate their perceptions of inequality into concrete behavior. Our findings show that this translation is never direct — it is always mediated by the social environment players inhabit. The next two themes explore how.

\subsubsection{Theme 3 Alliance Social Dynamics Shape Whether Players Respond Cooperatively or Aggressively}

The same perceived inequality produces different behavioral responses depending on the social environment a player inhabits. Alliance culture sets the behavioral norms, but social capital — the network of relationships and mutual support within that culture — determines what resources the system makes available to each player.

\textit{Two Alliance Models: Hierarchical Studios vs. Egalitarian Communities};  Players sorted into two distinct organizational cultures that shaped not just rules but fundamental behavioral patterns and even whale conduct. \textbf{Hierarchical "studio" alliances} which is organized less like a community and more like a competitive business. P03 described the most extreme version: "\textit{The top alliance is different. The people in power there are actually game studios, not ordinary players. They impose higher demands on members, concentrate resources on core players, and remove those who are inactive or disobedient.}" These professionalized alliances operated as businesses optimizing outputs. Rules were strict, questioning was prohibited, and members served as inputs rather than community participants. P09 described the behavioral atmosphere in her previous hierarchical alliance: "\textit{The hierarchy was very strict. Members couldn't question or oppose management. It felt very oppressive. Those oppressive alliances are usually much more aggressive and war-like.}"  In these environments, power gaps were not just visible, they were actively enforced through exclusion, resource concentration, and the suppression of dissent.

The second culture type was the\textbf{ egalitarian community alliance}, organized around shared participation rather than optimized output. P09 contrasted her current alliance: "\textit{My current one is very casual and peaceful ; anyone who wants to help manage can do so and be recognized. It's quite equal.}" P03 described her alliance's decision-making: "\textit{Our alliance is generally quite egalitarian. For some key decisions—such as whether to continue fighting or shift to peaceful farming—the alliance leader asks for everyone's opinions.}" The same power gaps existed in these alliances — the same spending differentials, the same ranking systems — but the cultural framing transformed how those gaps were experienced day to day.

\textit{How Culture Changes Whale Behavior; }Critically, alliance culture shaped not just members behaved but how \textit{whale behavioral norms}—the same high-spending players might act exclusively or inclusively depending on cultural context. P09 observed this directly: "\textit{In my old, strict alliance, the \#1 player only did activities with other top players. But in my current alliance, our top players are very easy-going and hang out with everyone.}" This suggested whale behavior wasn't purely individual personality but was culturally conditioned. Hierarchical alliances normalized exclusive social circles where whales only interacted with other whales, reinforcing status distance. Egalitarian alliances normalized inclusive interaction where whales participated in general chat and casual activities.

The behavioral consequences were stark. In hierarchical environments, players responded to power gaps with aggression that served psychological rather than strategic purposes. P02, embedded in a hierarchical environment where he occupied a subordinate position, described his response to being overpowered: \textit{"Even if I lose, I'll make them lose too. I don't care about my own loss. I want them to feel pain too... It's worth it for me because it makes me feel better emotionally."} These suicide attacks, which cost the attackers their lives and achieved almost nothing, came from a culture built on power and obedience, with no healthy way for people to deal with their anger and frustration. P09 described her previous alliance's broader pattern: \textit{"They often band together with 'Game Shills' to attack innocent players... This was hard for us 'peaceful' players to accept."} However, in egalitarian environments, the same competitive pressure produced collective rather than individual responses. P08 described his alliance's approach: \textit{"Because socializing and having fun as a team is what matters most. Fighting is just another experience. If the team decides to do it, I'll do it... If the whole alliance gets wiped out and loses all resources because the team isn't strong enough, there's no real loss."} Losses were reframed as shared experiences rather than personal defeats.

Beyond the alliance’s shared cultural norms, social capital—the web of relationships and mutual support within that culture—acted as a kind of community infrastructure. It shaped which resources the system made available to players when they faced setbacks. This infrastructure operated through three concrete mechanisms. First, it provided \textbf{material assistance}: resources, event access, and direct support after losses. P05 described how social activity translated into tangible benefits: \textit{"My social activity made leaders happy and helped me secure quota spots in important events. Even if I can't compete in power, I can contribute through organizing."} P06 observed the general pattern: \textit{"As long as you follow the alliance and talk actively, the distribution of resources and fortress rewards is quite fair."} Social labor — visible participation, helping with coordination, active communication — functioned as a currency exchangeable for material benefits, creating an achievement pathway independent of spending. Second, it provided \textbf{psychological reframing}: the same objective loss produced different outcomes depending on whether the social infrastructure was present to absorb it. P01 explained: \textit{"Some players, after their troops or resources are suddenly plundered and emptied, mistakenly believe they cannot recover and choose to quit the game. I think this is often related to social ties: if you know people and know who to ask for help, resource shortages can usually be alleviated. But if you don't spend money and lack social support, it's easy to interpret resource scarcity as 'the end of the road,' which is a heavy psychological blow."} The network did not change material reality — both the networked and isolated player lost the same resources. It changed what that loss meant. Third, it provided \textbf{meaning transformation}: shifting a player's frame from individual competition to collective purpose. P01 described this shift: \textit{"Once you join the social circle, the experience changes completely; you feel like you are fighting for a group of people rather than just for the game mechanics."} P03 articulated the same reorientation: \textit{"I gradually felt that playing together with my own allies was more important and that I shouldn't give up so easily."} When the community became the point, competitive losses stopped defining the experience.

\textit{The Social Capital Failure Player's Isolation;} P02's responses contained no references to alliance friendships, social activities, or relationship-based value. His description of game-play focused entirely on competitive frustration: "\textit{I've spent money myself, but seeing others spend tens of thousands, I feel very uncomfortable... Their single attack is stronger than my entire army.}" His escalating resentment—engaging in "suicidal attacks" for temporary emotional relief—suggested that without social capital, inequality became psychologically intolerable yet inescapable.

Together, alliance culture and social capital reveal that behavioral responses to inequality are never purely individual reactions to objective conditions. They are shaped by the social world a player inhabits — the norms that world enforces, the support it provides, and the meaning it makes available. The same gap, experienced through different social infrastructure, produces cooperation or aggression, resilience or collapse. What players actually do in response — the concrete behavioral repertoire that emerges from these conditions — is what Theme 4 examines next.

\subsubsection{Theme 4 Reported Game-play Behaviors in Response to Perceived Inequality}

Across our participants, behavioral responses to perceived inequality fell along a recognizable spectrum. Players who understood the game's economic logic and were socially embedded tended toward calculated, sustainable strategies. Players who lacked both understanding and social support tended toward behaviors that served emotional rather than strategic purposes. We describe these responses in turn.

The most common adaptive response was \textbf{strategic optimization} — players who fully understood the game's economic logic used that understanding to set rational limits on their own participation. Rather than competing endlessly against unreachable tiers, they identified a sustainable position and focused their effort there. P04 described what he called "retirement mode": once he reached Rally Leader threshold, he stopped pursuing further growth and maintained his position with minimal effort. He leveraged the alliance's structural need for Rally Leaders to maximize his status-to-effort ratio without escalating spending. P09 articulated the same logic: \textit{"The gap is a bottomless pit. Someone will always spend more. So you stop spending and just focus on daily activities to get the best score possible."} P03 similarly reduced her spending once she recognized the pattern of engineered obsolescence: \textit{"Now that events have become too repetitive, I've significantly reduced my spending."} These players did not quit. They identified a position they could sustain and organized their participation around it.

A second recognizable response was \textbf{selective withdrawal and alliance migration}. When the power gap within an alliance became too large or the alliance culture too oppressive, many players migrated rather than continuing to engage on unfavorable terms. P07 migrated twice. She first left her original server after her alliance was repeatedly overpowered by dominant whales. After migrating and becoming part of the strongest alliance in a new server, she migrated again when her own alliance imposed rules she found intolerable. P03 observed how widespread this pattern was in her server: \textit{"After we went through several large-scale conflicts, about half of the players in the server either quit the game or migrated. This shows that the top alliance mainly wants everyone to obey its arrangements, but about 50\% of players actually want to resist these arrangements. They just don't have the ability to do so, so they choose to leave."} For some players, migration was a recurring response rather than a one-time decision.

A third response was \textbf{social hedging} — players who could not compete economically redirected their effort toward building social relationships and organizational value within the alliance. P05 described her approach directly: \textit{"My social activity made leaders happy and helped me secure quota spots in important events. Even if I can't compete in power, I can contribute through organizing."} Rather than accepting marginal status, these players actively constructed alternative forms of contribution that the alliance genuinely needed. Social hedging was not a consolation strategy. It was a deliberate reorientation toward a parallel achievement pathway that the system made available. Players who pursued it successfully converted social labor into material benefits, status recognition, and a sense of agency that spending alone could not have provided.

The least adaptive response — and the endpoint of the spectrum — was \textbf{compensatory aggression}. This appeared specifically among players who perceived inequality as illegitimate but had no social infrastructure to absorb or reframe it. P02 describes this pattern most clearly. Facing a persistent and unbridgeable power gap, and without meaningful alliance relationships to provide support or alternative meaning, he responded by attacking stronger players knowing he would lose: \textit{"Even if I lose, I'll make them lose too. I don't care about my own loss. I want them to feel pain too... It's worth it for me because it makes me feel better emotionally."} Unlike strategic optimization or social hedging, compensatory aggression produced no material benefit and worsened his resource position. He continued because the attacks provided temporary emotional relief. Over time his frustration escalated rather than resolved, with no sustainable endpoint in sight.

Together these four responses reveal something important about how inequality functions in persistent online environments. Players are not passive recipients of structural conditions — they actively navigate them, selecting from a repertoire of responses shaped by what they understand, who they know, and what the system makes available to them. The spectrum from strategic optimization to compensatory aggression is not random. It maps directly onto the conditions Theme 3 identified: players with transparency, social embeddedness, and cultural support trend toward the adaptive end; players without these resources trend toward the destructive end. Understanding inequality in these environments therefore requires attending not just to the gap itself but to the full social and informational context in which players encounter it. 

\begin{figure}
    \centering
    \includegraphics[width=1\linewidth]{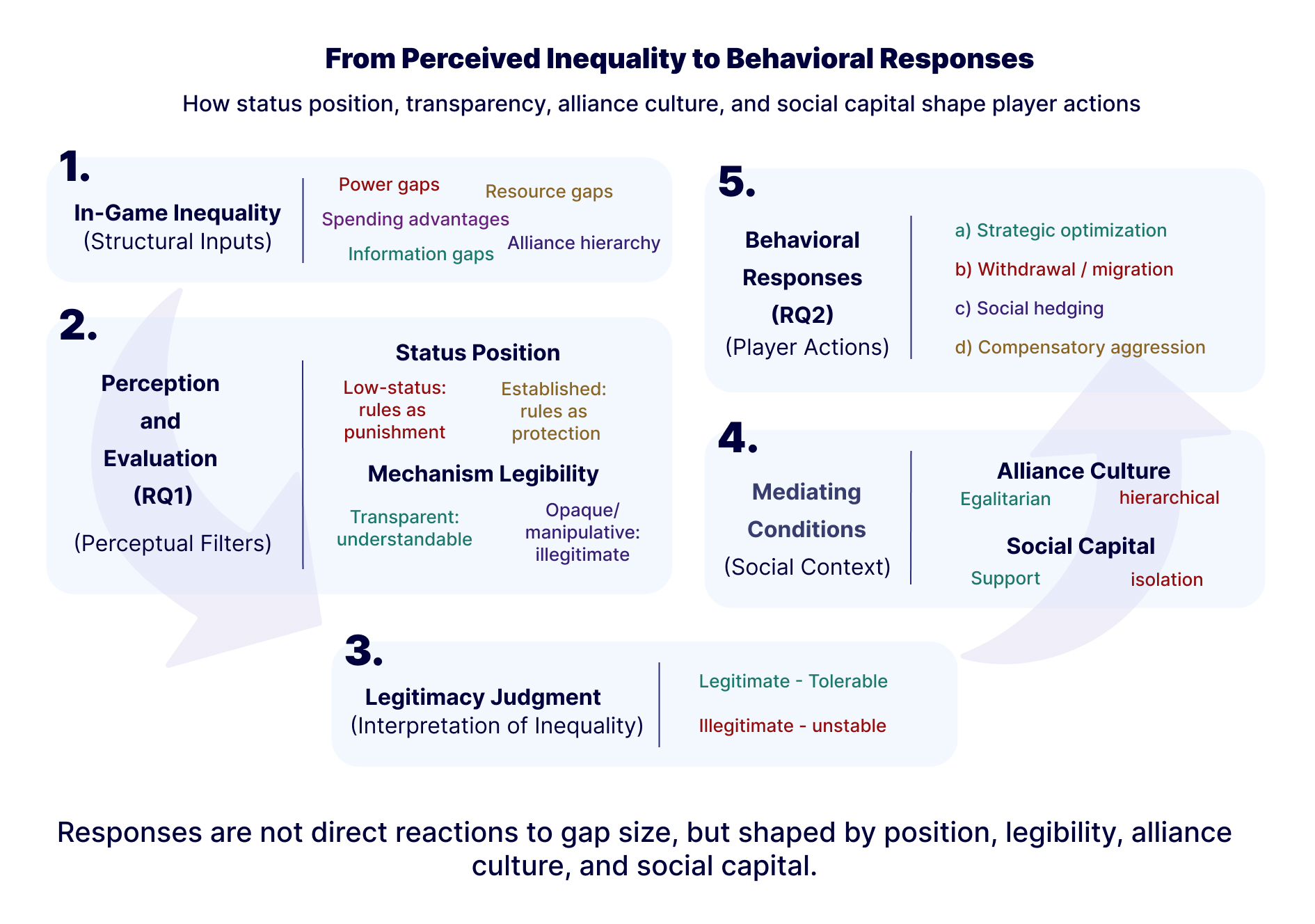}
    \caption{ An integrated model of how players perceive and respond to inequality in Whiteout Survival, showing the process from structural inputs through perceptual filters, legitimacy judgment, and mediating conditions to behavioral responses.}
    \label{fig:placeholder}
\end{figure}

\subsection{How is Inequality Maintained in the Community}

\subsubsection{Theme 5: Veteran On-boarding, Collective Enforcement, and Social Retention Reproduce the Hierarchy
}
The hierarchy in Whiteout Survival sustains itself not through rules imposed from above but through a socialization process embedded in the alliance social network — one that newcomers experience as community support but that gradually converts them into active participants in the hierarchy's reproduction.

New players arrive without access to the game's most important operational knowledge. Event strategies, diplomatic norms, resource management practices, and unwritten rules of conduct exist nowhere in the official game interface — they circulate in private WeChat groups organized by veteran players. To function effectively, newcomers must be recruited into these circles. P01 described how this works: \textit{"From the start, there are 'veterans' who come from older servers and know the ropes; they proactively create WeChat groups and recruit people. If you don't participate in these social circles, the game feels like a single-player experience, and it's hard to truly integrate into an alliance."} Veterans recruit newcomers, teach them the rules, and provide early material support. Nobody designed this onboarding system. Everyone maintains it.

What this process transfers is more than practical knowledge. When a veteran explains which rules matter and why, they simultaneously transfer their own framework for evaluating those rules — shaped by their position in the hierarchy and their investment in its stability. P01 described the shift that follows full integration: \textit{"Once you join the social circle, the experience changes completely; you feel like you are fighting for a group of people rather than just for the game mechanics."} The game stops feeling like a competitive system that distributes resources unequally and starts feeling like a community worth protecting. The hierarchy that produces the inequality becomes the hierarchy that protects the community.

P06's trajectory shows how complete this transformation can become. Six months after entering the game as an angry critic of its power structures, he was actively enforcing the same rules that had once punished him: \textit{"We all worked together against them. Even if we couldn't win 1-on-1, we would all join rallies to hit them during alliance events. Eventually, they couldn't participate in any activities because the whole server blocked them, so they had to leave."} He was not instructed to do this. He did it because the rules now protected something he valued. The hierarchy reproduced itself through an act he experienced as loyalty.

Experienced players frequently recognized that the game's inequality was deliberately engineered. P03 identified the psychological mechanics behind escalating spending: \textit{"Many mechanics are designed so that you're 'just a little short' of getting the reward, which pushes you to keep spending. This is a very typical design."} P09 identified developer-planted provocateurs designed to manufacture competitive conflict: \textit{"They use 'Shills' to provoke Whales into fighting... They intentionally design these unequal mechanisms to maximize profit."} Yet this recognition did not produce outrage or exit. P09 explained: \textit{"After playing for a few days, you are mentally prepared for the fact that it's unfair. But you also know it's a selling point to attract Whales. So, you accept it."} Veterans passed this adapted worldview on to newer players as part of the socialization content itself — teaching them not just the rules but how to make peace with the system behind them. Recognition becomes another mechanism of reproduction rather than a challenge to it.

The community that remains after this process is also shaped by who has already left. As established in Theme 2, players who could not accept the game's transparent economic hierarchy exited early and decisively. The community a newcomer enters is therefore already composed predominantly of players who have accepted or been socialized into accepting the hierarchy's terms. Dissenting frameworks are rare. The collective investment in the existing structure is substantial. This makes each new player's socialization faster and more complete — there are few alternative voices left to offer a different way of reading the system.

Social ties also created retention that operated independently of whether players still enjoyed the game's content. P04 observed: \textit{"Many people keep playing not because of the game content itself, but because they do not want to lose their social circle."} P03 described her own experience: even after recognizing the pattern of engineered obsolescence and significantly reducing her spending, she continued playing because \textit{"playing together with my own allies was more important."} The relationships players built over months of daily interaction became a reason to stay that had nothing to do with the game's competitive systems. Leaving meant abandoning a social world that had become genuinely meaningful. It's a cost that rose with every month of deeper integration.

Taken together, these mechanisms — informal veteran-led onboarding, evaluative framework transfer, collective enforcement, recognition without resistance, self-selection, and social retention — describe a hierarchy that reproduces itself through the social infrastructure players experience as belonging and support. No single player intends this outcome. It emerges from the accumulated effect of small, ordinary acts: teaching a newcomer the rules, pulling someone into a WeChat group, enforcing a norm against a rule-breaker, staying another month because of friends. The result is a system that feels, from the inside, like a community worth protecting.

\subsubsection{Theme 6: Ecosystem Interdependence, Alternative Achievement Pathways, and Social Participation Sustain Willing Participation Under Inequality}

The foundation of this willing participation is mutual dependency. The game's economy only works because whales and free-to-play players need each other. P03 articulated this directly: \textit{"If there are no free-to-play players, even the big spenders will feel bored, and the whole ecosystem will decline."} Whales need F2P players as opponents to defeat, audiences to witness their achievements, and content to populate events. F2P players need whales as Rally Leaders who enable high event scores, as protectors during conflicts, and as resource generators whose alliance spending benefits everyone. P06 described the concrete value of having a strong whale in the alliance: \textit{"If there's a car head [Rally Leader] in the alliance, other members actually benefit. On the one hand, it's easier to win fights; on the other hand, you can obtain more resources during events."} P03 observed that developers actively managed this balance: \textit{"If the numerical gap becomes too large and there's too little room for strategy, players will leave. So the developers strengthen some free-to-play heroes to retain non-spending players."} The inequality that looks purely extractive from the outside creates genuine interdependence from the inside. Neither side of the power gap can have a functioning game without the other.

Beyond interdependence, the hierarchy sustained engagement because it supported multiple parallel definitions of what it meant to succeed. Spending was one route to status — but not the only one. P09 observed two distinct status hierarchies operating simultaneously: \textit{"Those with high status are either Whales with raw power or people with strong social skills. The social types handle management."} Alliances needed both economic power and organizational competence — and they recognized both. This created real upward mobility for players who were willing to invest social and organizational labor rather than money.

Players pursued this pathway in concrete, recognizable ways. P07 entered a top alliance not through spending but through management skills: \textit{"I was a manager in a small alliance, and because they needed management talent and someone with decent power, they invited me."} P05 reconstructed her position within the alliance around organizational contribution: \textit{"Being a useful Filler is actually quite meaningful. I actively participate in alliance activities, help with event registration, communicate with members, and even though I can't contribute as much combat power, I can contribute through organizing and enthusiasm."} Her social activity translated directly into material benefits — secured quota spots in important events, resource allocation, and status recognition that her combat power alone would never have earned him. The hierarchy that appeared to privilege only spending actually valued organizational competence. Non-whales who invested social labor could achieve genuine recognition within it.

P04's experience illustrated a different form of agency — strategic restraint. After reaching Rally Leader threshold, he stopped pursuing further growth and maintained his position with minimal effort: rather than escalating spending to chase constantly moving goalposts, he identified a sustainable position and stayed there. He leveraged the alliance's structural need for Rally Leaders — alliances needed him more than he needed any particular alliance — to maximize his status relative to his effort. This was not passive acceptance of limitation. It was an active, calculated choice about how to engage with the system on favorable terms.

For some players, the competitive hierarchy was almost beside the point. P08 played primarily to socialize: \textit{"I play purely to chat; my work is busy, so I just click around during my commute. To me, socializing is more important than power or skill."} He participated in collective alliance actions purely for the experience of doing things together: \textit{"If the team decides to do it, I'll do it... If everyone is charging forward together, I'll join in."} Even total alliance defeat was acceptable because social solidarity was the actual source of value: \textit{"There's no real loss."} For P08 the game's inequality structures were background noise. The social world the game hosted was what he was there for — and that social world was available to him regardless of how much he spent.

All in all, these players — the organizational climber, the strategic retiree, the pure socializer — describe a system that sustains participation not by trapping people but by offering genuine value across different definitions of what the game is for. Willing participation in an unequal system is not simply the result of successful socialization. It reflects real agency operating within real constraints — players finding workable terms of engagement and choosing, on those terms, to stay.

\begin{figure}
    \centering
    \includegraphics[width=1\linewidth]{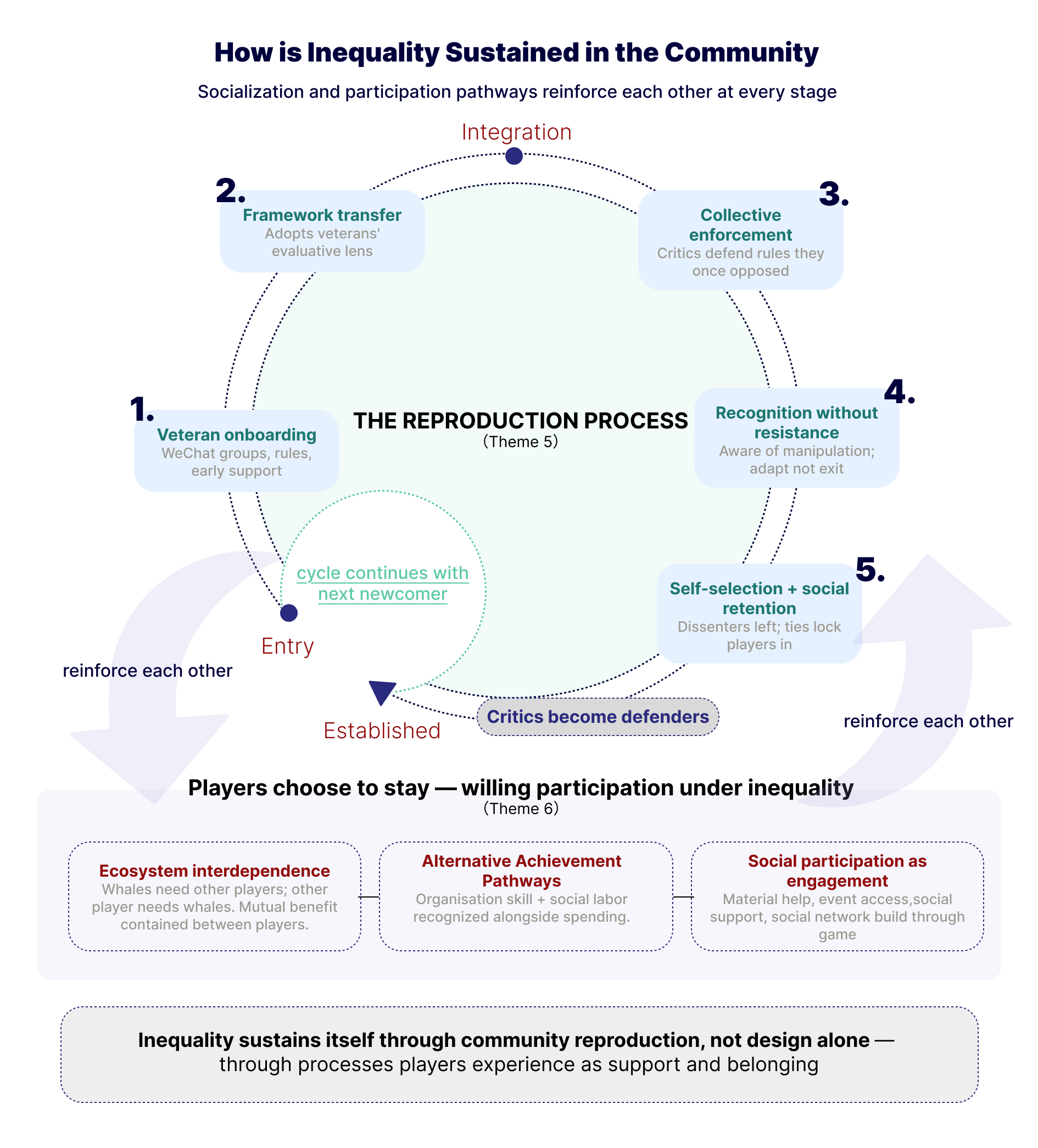}
    \caption{ System diagram shows inequality sustained in the community. The open cycle illustrates the socialization process; the three boxes below show parallel participation pathways that sustain willing engagement.}
    \label{fig:placeholder}
\end{figure}

\section{Discussion}\label{sec:Discussion}
Our findings reveal how inequality operates in digital gaming environments through mechanisms that are simultaneously familiar to social psychology and distinctly shaped by the affordances of persistent online worlds. We organize the discussion around four theoretical moves: extending Payne's Broken Ladder framework into digital contexts; reconceptualizing transparency as the primary legitimacy mechanism in monetized game communities; establishing social capital as adaptive infrastructure for sustained engagement under inequality; and identifying design implications for governing hierarchical systems without dismantling them.

\subsection{Position-Dependent Legitimacy: Extending and Complicating Payne in Digital Contexts}

Payne's Broken Ladder framework argues that perceived inequality — subjective sense of relative position rather than objective resource gaps — exerts more powerful psychological and behavioral effects \cite{payne2017economic}. Our findings strongly support this proposition in digital gaming contexts. Players who initially condemned structural disparities frequently defended the same systems after accumulating resources or gaining social integration, often within weeks. These transformations were driven by subjective positioning, not by changes in the game's underlying mechanics.

However, extending Payne into digital environments reveals important boundary conditions. First, Payne models perceived disadvantage as a relatively stable trigger of stress and defensive adaptation, yet our findings show that digital mobility — the ability to migrate servers, switch alliances, and accumulate resources rapidly — transforms inequality from a chronic structural burden into an episodic one. Where Payne's subjects face relatively fixed hierarchies, Whiteout Survival players could reposition themselves within weeks, compressing what would take years in material contexts. This temporal acceleration means that the psychological effects Payne associates with long-term disadvantage appear and dissolve far more rapidly in digital environments, suggesting that chronic versus acute stress exposure may be a meaningful moderator of the Broken Ladder effect that future work should examine.

Second, Payne predicts that perceived inequality systematically erodes institutional trust \cite{payne2017economic}. Yet our study shows that trust is not uniformly undermined when structural interdependence exists. Because whales, strategic leaders, and free-to-play players rely on one another for event performance and ecosystem viability, players reframed extreme disparities through cooperative logics rather than zero-sum competition. This finding connects to game research on how game interdependence structures social outcomes: Depping et al. \cite{depping2018designing} demonstrate that interdependence, not cooperation, is the primary predictor of social capital formation in multiplayer games, a finding our study extends by showing that structural interdependence also moderates legitimacy judgments under inequality conditions.

Third, and most critically, our data reveal that the Broken Ladder effect operates through identity as much as through cognition. Payne's framework treats legitimacy shifts as primarily cognitive,  individuals recalculate their relative position and update their attitudes accordingly. But P06's transformation from angry victim to rule-enforcer, and P07's group-level shift from oppressed migrants to inequality-imposing elites, suggest something deeper: players do not simply change their minds about inequality; they change who they are within the community. This connects to game research on ranking and distinction: Kou et al. \cite{kou2016ranking} show that ranking systems in competitive games like League of Legends are not merely informational but constitutive of player identity, shaping how players narrate their trajectories and distinguish themselves from others. Our findings extend this work by demonstrating that such identity formation operates across the full social hierarchy, not just through skill-based rank, and that identity shifts in response to stake acquisition can reverse players' ideological positions entirely — producing the "dragon slayer" pattern where yesterday's critics become tomorrow's enforcers.

\subsection{Exit as Adaptation Rather Than Acceptance}

The "Broken Ladder" framework proposed by Payne \cite{payne2017economic} is grounded in real-world social inequality, with the implicit premise that individuals often find it difficult to break free from the social structures in which they are embedded. When inequalities in income, education, occupational status, or social class persist over the long term and opportunities for social mobility are scarce, individuals can only adjust their cognition and behavior within the existing structure to cope with a persistent sense of relative deprivation. However, this study finds that inequality in digital gaming environments is characterized by a significant and unique boundary condition: players always retain the possibility of exiting the system. Theoretically, they can stop playing, switch serves, change alliances, or even reduce their level of participation. Therefore, compared to real-world society, inequality in games is not an entirely inescapable structural predicament, but rather the "acceptability system of inequality".

This discrepancy suggests that players’ continued participation in the system cannot simply be interpreted as an endorsement or acceptance of inequality. Like P3 said "Some alliances were simply disbanded, and players who spent little money were kicked out. Eventually, I came to realize that playing with my allies was more important, and I couldn’t just give up on them. After that, I played less and less". It shows that players are fully aware of the resource advantages enjoyed by paying players and the hierarchical structure within the alliance, yet they do not immediately leave the game as a result. Instead, they constantly weigh the options of “continuing to participate” against “exiting the system.” For players, the real question that needs to be addressed is not why they accept inequality, but why they choose to stay despite being fully aware of it.

This finding resonates with recent discussions on disengagement in game studies. Kosa et al.\cite{kosa2024disengagement} note that disengagement and withdrawal should not be viewed as failures of the player experience, but rather as integral components of players’ self-regulation of their experiences. Our study further suggests that, within the context of long-running SLG environments, this disengagement serves not only as a means of regulating the gaming experience but also as an adaptive strategy in response to structural inequalities. Rather than directly challenging the system itself, players reduce the psychological costs associated with inequality by decreasing their competitive investment, shifting toward social activities, changing alliances, or redefining their goals.

Furthermore, our findings expand upon Bergstrom’s\cite{bergstrom2019quitting} study of game exit processes. Bergstrom notes that exiting a game is not a binary shift from “playing” to “not playing,” but rather a process of ongoing negotiation and repositioning. Players in Whiteout Survival similarly rarely leave the game outright, but instead constantly shift between different forms of engagement. They may withdraw from competitive leagues while maintaining social connections, abandon leaderboard competition but continue participating in collective activities, or even temporarily leave and then return. In other words, they are not exiting the game itself, but rather renegotiating their relationship with its unequal structures.

Therefore, we argue that players’ continued participation should not be interpreted as mere acceptance of inequality, but rather as a form of “adaptive retention,” since the act of not leaving is itself an adaptive strategy. Players remain within the system not necessarily because they endorse the hierarchical structure, but perhaps because leaving would also entail the loss of accumulated social capital, identity, and a sense of community belonging. In this sense, the persistence of inequality is not based on complete legitimization, but rather on players’ ongoing negotiation of the costs of leaving, social dependencies, and potential benefits. Therefore, rather than interpreting player behavior as “acceptance” of inequality, our research suggests it should be understood as a form of “adaptive negotiation.”

Theoretically, when studying inequality in digital environments, it is necessary to incorporate the concept of “exit” into the analytical framework. While inequality in the real world is typically constrained by institutional and resource conditions, inequality in online games is characterized by a continuously available option to exit. Consequently, inequality in games differs both from the inescapable structural inequality of the real world and from short-term consumer relationships that can be abandoned at any time. It is an “exit-able but costly inequality system” that falls somewhere in between. Whether players choose to stay does not directly reflect their acceptance of inequality, but rather reflects how they continuously negotiate the trade-offs between inequality, social dependence, and the costs of leaving.

\subsection{Inequality Perception Beyond Game Balance}

Prior CHI PLAY research has examined pay-to-win fairness primarily through the lens of competitive balance, showing that acceptance of paid advantages depends on whether monetization disrupts core game-play \cite{freeman2022pay}. Our findings both confirm and significantly extend this work. We confirm that players do not uniformly reject paid advantages; many accept, and even embrace, economic hierarchy. However, the determinant of acceptance in our study is not whether balance is disrupted — it is whether the mechanisms producing inequality are visible and predictable.

This distinction matters theoretically. Freeman et al. \cite{freeman2022pay}analyze fairness as a judgment about outcome equity: did paid advantages produce unfair results? Our participants' reasoning reveals a prior evaluative stage: do I understand how this system works? P04's "upward severing" — psychological disconnection from unreachable tiers once their cost became fully transparent — illustrates that transparency can enable acceptance of extreme inequality that would be intolerable under opacity. Players who understood that "Twin Dragon Overlord" status required sustained spending of hundreds of thousands of USD did not aspire to it; they disengaged from it as a comparison target altogether. This is Festinger's Hypothesis V operating in action: comparison ceases when the gap becomes too large to provide useful self-information \cite{festinger1954theory}. Critically, transparency is what allows players to recognize when a gap is too large to bridge, enabling rational disengagement rather than sustained frustration.

We term this mechanism a "functional power contract": high-spending players' dominance is accepted as legitimate when it is visible, predictable, and tied to collective benefits. This extends procedural justice theory \cite{ariely2017corruption} into gaming contexts by showing that fairness judgments in highly unequal digital systems are shaped more strongly by the intelligibility of rule structures than by the equality of outcomes. Legitimate inequality is legible inequality.

This mechanism has an important counterpart. When the same structural inequality becomes opaque — through gate-kept information channels (P01's WeChat group problem), revealed hypocrisy (R07's dragon slayer experience), or deliberate manipulation (P09's developer shills) — legitimacy collapses regardless of the magnitude of the gap. This mirrors CHI PLAY findings on how the game moderation environment shapes player trust: research on toxicity and community governance shows that players' tolerance for negative experiences depends heavily on whether governance processes are transparent and consistent \cite{frommel2023how}. In our context, the relevant governance is not moderation of behavior but transparency of monetization logic — yet the underlying mechanism is the same. Perceived arbitrariness, whether in how rule-breakers are punished or how power advantages are constructed, destroys legitimacy and provokes reactive disengagement.

Although our study finds that transparency plays a significant role in shaping players’ acceptance of inequality, this does not mean that transparency alone is sufficient to automatically legitimize inequality. In other words, while transparency is an important prerequisite for players’ acceptance of inequality, it is not a sufficient condition. As P01 said "Resource disparities alone do not significantly influence my view of inequality. I tend to see such disparities as the result of individual choices rather than systemic bias favoring certain players. As long as the mechanisms are transparent, I can generally accept this kind of resource inequality.". Which means players are often able to tolerate or even rationalize highly unequal resource gaps, but this acceptance is contingent upon the simultaneous fulfillment of multiple conditions, not merely the visibility of the rules.

This finding expands upon existing research on procedural justice and game fairness. Freeman et al. \cite{freeman2022pay} note that players’ acceptance of pay-to-win advantages depends on whether they disrupt the core game balance. Procedural justice highlights that people’s evaluations of outcomes are often influenced by whether the decision-making process is transparent, fair, and consistent \cite{ariely2017corruption}. Our research supports this view, but we further found that in highly monetized, long-running SLG environments, transparency does not primarily serve to make inequality fair, but rather to make it intelligible. When players understand how power imbalances arise, what costs are involved, and where they stand in the hierarchy, they are better able to form stable expectations about these imbalances and adjust their points of comparison, participation goals, and engagement strategies accordingly.

However, our interview also shows that the effectiveness of transparency has clear limits. When players perceive that, although the rules are transparent, their implementation runs counter to the collective interest, undermines the ecosystem of the alliance, or involves manipulation, legitimacy can still collapse rapidly. For example, the “dragon slayer” experience in P07, P09’s description of developers manipulating competition, and the information blackout experienced by P01 all occurred after players had already understood the system’s operational logic. It was not a lack of information that caused dissatisfaction, but rather their belief that this power structure no longer served a common goal, having instead become a tool for unilateral profit-taking. This suggests that while transparency can reduce uncertainty, it cannot eliminate normative judgment. Players are concerned not only with “how inequality operates,” but also with “why inequality exists” and “who benefits from it.”

The implications of these findings for understanding pay-to-win fairness are substantial. Existing studies primarily focus on whether players accept inequalities in outcomes, such as who wins and who loses. Our findings suggest that an equally important question concerns how players evaluate the processes through which those inequalities are produced, maintained, and justified. Transparency helps players understand how unequal systems operate, but understanding does not automatically translate into acceptance. Instead, legitimacy emerges through the interaction of transparency, perceived procedural fairness, collective benefit, and social interdependence.

This suggests that players’ tolerance for outcome inequality and their tolerance for procedural opacity may represent analytically distinct dimensions shaped by different mechanisms. A system may be highly unequal yet perceived as legitimate when its rules are understandable, predictable, and embedded within mutually beneficial social structures. Conversely, even transparent inequalities may become unacceptable when players perceive manipulation, arbitrariness, or the breakdown of collective interests. Future research should therefore move beyond asking whether players perceive inequality and examine the social and institutional conditions under which inequality becomes legitimate, contested, or rejected.

\subsection{Spending as a Long-Term Social Commitment Rather Than a Purchase Decision}

Our research further demonstrates that in Whiteout Survival, high spending is not a one-off purchase, but rather a long-term, gradual investment process embedded in the social structure. The thousands of dollars in spending reported by respondents often accumulate over months or even longer through event bundles, hero iterations, ranked competition, and alliance activities. More importantly, this spending is not an individual choice independent of the game's social environment, but is closely related to maintaining relative status, fulfilling alliance obligations, and preserving group identity.

This finding expands on existing research on paid behavior and predatory monetization. Previous studies typically understand consumption as an immediate reaction of players to specific monetization incentives, such as scarcity design, limited-time events, or competitive pressure\cite{petrovskaya2022predatory, hamari2017}. However, our data suggest that in continuously operating strategy game (SLG) environments, consumption is closer to a sustained social commitment than a single transaction. Players are not constantly pursuing absolute dominance, but rather continuously investing resources to maintain their existing position, avoid falling behind, and meet role expectations in alliance competition.

This perspective also helps explain why many players choose to continue participating even when they recognize the inequalities in the system. As described in P04, players eventually realize that what they are pursuing is not constantly improving their strength, but avoiding falling out of their existing tier. Consumption thus gradually shifts from a tool for gaining an advantage to a cost of maintaining status. This means that inequality in the game is not created through single-time consumption, but is continuously produced and reproduced through long-term competition, social comparison, and repeated investment. For HCI and game studies, this suggests that we should understand economic inequality as a dynamic, cumulative process, rather than simply a question of whether players buy a particular paid item.

\subsection{Social Capital as Adaptive Infrastructure: The Buffer in hierarchical community}

Social capital emerged as the most pervasive adaptation mechanism in our study, with 10 of 11 focal participants describing reliance on social ties to navigate inequality. The sole participant without meaningful social integration (P02) was also the only participant exhibiting consistent failed adaptation — escalating resentment, compensatory aggression, and inability to reframe losses. This near-universal pattern suggests that social capital functions not merely as a coping resource but as adaptive infrastructure for sustained engagement under extreme inequality conditions.

This finding connects to and extends a growing body of HCI research on social capital in online gaming. Depping et al. \cite{depping2018designing} demonstrate that in-game social capital is strongly associated with reduced loneliness and improved relatedness satisfaction among multiplayer game players, identifying interdependence as the primary game design predictor of social capital formation. Our findings extend this work in two directions. First, we show that social capital operates specifically under inequality conditions — its buffering function is not simply about general well-being enhancement but about preventing catastrophic loss interpretation. The same objective event (being "zeroed") produced radically different outcomes for socially integrated versus isolated players. Networked players treated resource loss as a temporary setback; isolated players interpreted it as structural defeat. Social capital did not change material reality; it transformed how inequality was experienced. Second, Depping et al. examine social capital as an outcome of game design properties, while our study reveals it as a behavioral strategy that players actively cultivate in response to perceived inequality — joining WeChat groups, participating in alliance events, performing visible social labor. This active, strategic dimension of social capital under inequality has not been previously examined in the HCI games literature.

This finding also resonates with research on how community toxicity undermines social connectedness. Frommel et al. \cite{frommel2023how} show that perceived community toxicity is associated with lower in-game social capital, reduced relatedness satisfaction, and increased loneliness. Our hierarchical alliance data mirrors this pattern: the most aggressive, war-like alliances described by P09 and P02 produced precisely the social isolation and compensatory behavior that Frommel et al. associate with toxic community environments. Conversely, the egalitarian alliances described by P08 enabled the social capital accumulation that buffered inequality's effects. This suggests that alliance culture — the behavioral norms governing how power is exercised and how members interact — operates as a mediating variable between structural inequality and social capital, a relationship that future quantitative work could productively examine.

Social capital in our study also created what we term alternative achievement pathways — status routes independent of economic investment. P07's entry into a top alliance through management skills, R05's use of organizational labor to secure event access, and P08's purely social participation all illustrate how non-whale players navigated hierarchy through social-organizational competence rather than spending. This dual status hierarchy — economic power coexisting with social capital as a parallel route to recognition — has a counterpart in previous research on role structures in online games. Xie et al. \cite{xie2022roleseer} demonstrate that MMORPGs support heterogeneous informal social roles that coexist with formal hierarchies, and that these informal roles are fluid and transitional rather than fixed. Our findings extend this work into an economic inequality context: the social roles occupied by alliance managers, event coordinators, and community builders function as a structural counterweight to the spending-based formal hierarchy, enabling participation and status for players who cannot compete economically.

This study further reveals a state of ambivalent adaptation among non-paying players in a highly unequal gaming environment. Unlike the simplistic view of non-paying players as either victims or supporters of the system, our interviews reveal that they often occupy a position where criticism and dependence, resistance and participation coexist. On the one hand, nearly all interviewees clearly recognized the resource gap, competitive disadvantage, and structural unfairness caused by the advantages of paying players. They frequently expressed dissatisfaction with the dominance of “whale players,” pay-to-win mechanisms, and the power structures within the alliance. On the other hand, these players continue to rely on the very same unequal structures to obtain alliance protection, event rewards, a sense of social belonging, and a sense of alternative achievement. Therefore, they do not fully accept inequality, but rather continuously adapt to it.

This finding expands upon existing research on social capital and online gaming participation. While Depping et al. \cite{depping2018designing} primarily emphasized the crucial role of social capital in enhancing a sense of relatedness and reducing loneliness, our study further demonstrates that, in highly unequal SLG environments, social capital serves not only as a resource for social support but also as a vital adaptive mechanism for maintaining player retention. The reason non-paying players are able to continue participating is not because economic inequality has disappeared, but because alliance relationships, shared goals, and social identity partially offset the psychological impact of that inequality. In other words, social capital does not alter the objective position in which players find themselves, but it does change the way they perceive that position.

Compared to Payne's view that individuals' responses to inequality depend on their perceived subjective position rather than objective resource gaps \cite{payne2017economic}, our research further demonstrates that even when players consistently perceive themselves to be in a lower position, they may still derive alternative identity value through social roles such as alliance members, event organizers, and coordinators, thereby redefining their standards of success. This adaptation is not an acceptance of inequality, but rather a strategic adjustment designed to reduce the psychological costs of inequality. As demonstrated by P07, P08, and P05, players do not deny their inability to compete with whales at the consumption level, rather, they proactively shift toward pathways such as organizational coordination, community building, and social contribution to maintain their sense of self-worth.

The broader implication for HCI is that social capital should not be understood merely as a desirable byproduct of multiplayer game design, but as a critical infrastructure that enables players to remain engaged despite persistent structural inequality. Our findings suggest that many non-paying players continue participating not because they perceive the system as fair, but because social relationships, collective identities, and alternative achievement pathways make unequal conditions psychologically manageable. In this sense, social capital does not eliminate inequality, it allows players to negotiate, reinterpret, and adapt to it.

This perspective extends prior HCI research by showing that the sustainability of unequal digital communities depends not only on economic incentives or progression systems, but also on the availability of social mechanisms that buffer dissatisfaction and support continued participation. Consequently, systems that undermine social capital formation, through opaque governance, exclusionary alliance cultures, or insufficient coordination infrastructure, may do more than reduce player enjoyment. They may remove the very resources through which non-paying players cope with inequality, making structural disparities psychologically intolerable.

\subsection{Inequality as Predatory Design: Connecting to the Dark Patterns Debate}

Our findings contribute to an ongoing debate in CHI PLAY and HCI about whether the structural conditions that produce player inequality in monetized games are incidental features of competitive design or deliberate instruments of revenue extraction. The evidence  from our participants points firmly toward the latter — and in doing so, extends the predatory monetization literature into a context it has not previously examined.

Prior work has cataloged the techniques through which predatory monetization operates. Petrovskaya and Zendle~\cite{petrovskaya2022predatory} identified 35 such techniques from the player perspective, including engineered competitive pressure, manufactured scarcity, and information asymmetry. Zhang et al.~\cite{zhang2025microtransactions} demonstrated how these techniques extend beyond individual financial harm into the social fabric of game-play, showing that monetization systems in social games leverage interpersonal relationships and community belonging as spending pressure — a phenomenon they term \textit{social monetization}.  Hamari et al.~\cite{hamari2017} established the underlying demand-side mechanism: artificial design obstacles that interrupt \textit{unobstructed play} are among the strongest predictors of actual in-game spending, suggesting that friction-by-design is a reliable monetization lever.

Our participants recognized precisely these mechanisms in Whiteout Survival — and named them with notable clarity. P03 described the near-miss escalation trap: \textit{"Many mechanics are designed so that you're just a little short' of getting the reward, which pushes you to keep spending. This is a very typical design."} P04 identified engineered obsolescence: each generation of heroes requires renewed spending to maintain relative position, ensuring that past investment continuously depreciates. P09 reported the most explicit manipulation — developer-planted provocateurs who manufacture competitive conflict to stimulate whale spending. These are not players speculating about developer intent; they are experienced participants articulating, in their own terms, the dark patterns literature's core taxonomy.

What our study adds to this literature is a finding about how these mechanisms interact with player perception over time. Crucially, recognition of dark patterns in our data did not produce the indignant rejection or exit behavior that the predatory monetization literature might predict. Instead, it enabled what we term \textit{informed disengagement}: players who could accurately diagnose manipulative mechanics used that knowledge to set rational spending limits, disengage from competitive escalation, and optimize for sustainable participation. This stands in contrast to the felt manipulation and regret documented by Zhang et al.\ in casual UGG contexts~\cite{zhang2025microtransactions}, and suggests that the psychological consequences of dark patterns are moderated by the persistence of the game environment and the depth of players' structural literacy. In transient, episodic games, dark patterns produce surprise and exploitation; in persistent SLG communities built around months of daily engagement, experienced players develop a meta-cognitive map of the monetization system that partially protects them from its intended effects.

This distinction has an important implication for how inequality functions as a monetization instrument. Hamari et al.~\cite{hamari2017} found that competitive motivation — the desire to outperform others — does not significantly predict in-game spending volume. Our findings help explain this counterintuitive result. In Whiteout Survival, the players who spent most aggressively were not primarily motivated by competitive dominance but by social obligation, identity investment, and the desire to provide collective benefits to their alliances. Inequality, in this reading, is not simply a pay-to-win competitive mechanism; it is a \textit{social architecture} that recruits social identity and interpersonal commitment as spending pressure. This represents a more sophisticated form of predatory design than the artificial obstacles and loot box mechanics that dominate the existing dark patterns taxonomy — one that operates through community structure rather than interface manipulation alone.

The implication for the CHI PLAY community is that the dark patterns debate may need to expand its analytic scope. Existing frameworks~\cite{petrovskaya2022predatory, zhang2025microtransactions} are well-suited to identifying techniques that exploit individual psychological vulnerabilities at the point of purchase. They are less equipped to analyze monetization systems that operate through the long-term engineering of social environments --- systems in which the \textit{community itself} becomes the mechanism of spending pressure. Future research and design advocacy should address not only what happens at the purchase interface but how persistent social architectures are designed to sustain and normalize inequality in ways that make exit psychologically costly and resistance structurally difficult.

\subsection{Design Insights: Governing Inequality in Digital Social Systems}

Our findings suggest that inequality in persistent SLG games is neither inherently harmful nor inherently sustainable. Its psychological consequences depend on three design-controllable conditions: the legibility of the mechanisms that produce it, the cost of exiting it, and the range of status pathways available within it. We translate each into a concrete design direction below.

\subsubsection{Display Spending Tier Indicators and Ecosystem Contribution Data at the Point of Comparison}

In Whiteout Survival, and similar SLG game, extreme structural inequality—particularly through pay-to-win mechanics, inequality between players is made highly visible through power scores, combat rankings, and progression milestones, yet the cost structures that produce these gaps are confined to store menus that players rarely consult at the moment of comparison. Critical coordination knowledge circulates through private external channels rather than in-game systems, concentrating informational power among veteran players and leaving newcomers structurally disadvantaged. This spatial mismatch means players encounter inequality constantly but lack the information needed to interpret it rationally.

Our findings show that players did not resist inequality because gaps were large. They resisted it when they could not understand where the gaps came from. When cost structures were transparent, players made rational decisions about which comparisons to pursue and which to abandon — a pattern consistent with Festinger's social comparison theory, which predicts that comparison ceases when a gap becomes too large to provide useful self-information, but only when that threshold is actually visible. Our data further show that players who understood the ecosystem's mutual dependency accepted extreme inequality far more readily. Players who grasped that whales need free-to-play members as opponents and audience, and that free-to-play players need whales for Rally Leadership and event performance, reframed the hierarchy as interdependence rather than exploitation. This understanding emerged slowly through experience, not through anything the game communicated directly. Petrovskaya and Zendle \cite{petrovskaya2022predatory} identify information asymmetry as a core predatory technique; our findings confirm that it operates specifically at the moment of social comparison, not at the purchase interface.

We therefore recommend surfacing approximate spending tier indicators directly on leader-boards and player profiles, so players can calibrate comparison targets at the moment those comparisons occur. We further recommend introducing an in-game unified communication channel for event strategies and alliance announcements, redistributing the informational power currently held by veterans. Finally, we recommend interface elements that make ecosystem interdependence visible early in a player's progression, showing new players what high-spending alliance members generate for collective events and showing high spenders the participation data their free-to-play members contribute. Transparency about cost enables rational disengagement from unreachable tiers. Transparency about interdependence enables acceptance of the tiers that remain.

\subsubsection{Implement Soft Migration Pathways and Pre-Join Alliance Culture Indicators}

In persistent SLG game communities, server migration is currently a permanent and socially costly decision. Players who leave an alliance lose their accumulated relationships, informal reputation, and community standing built over months of daily interaction. No pre-join information about alliance culture, governance style, or conflict patterns is available to players before they commit to a new community. Classical inequality research often interprets withdrawal as a symptom of alienation or distrust \cite{kaiser2009playerrating} . And game design has largely followed this framing by treating exit as a problem to be prevented rather than a behavior to be supported. However, our findings resonate with theoretical perspectives on coping and system navigation \cite{taylor1984theoretical, kay2011social}, showing that players frequently respond to untenable inequality through strategic attrition, selective engagement, or server migration rather than escalation. These behaviors do not necessarily reflect disengagement from the community, rather, they represent adaptive recalibration in contexts characterized by mobility and low exit costs.

Players in our study moved servers deliberately when power gaps or alliance cultures became untenable, yet current design makes migration permanent and socially costly. Kosa et al. \cite{kosa2024disengagement} argue that disengagement from games should be treated as a design problem in its own right: rather than focusing exclusively on engagement and retention, designers should support players in exiting in a self-determined way. This framing matters in our context because quitting in WOS is not a failure of engagement — it is often a rational response to structural conditions. Bergstrom \cite{bergstrom2019quitting} makes a similar point about persistent online games more broadly: players rarely quit permanently; they take breaks, return, migrate, and re-engage depending on social and contextual factors. The binary of "playing" versus "quit" obscures a much more fluid reality.

We therefore recommend introducing a time-limited soft migration feature, allowing players to trial a new server for a defined period while retaining read-only access to their original alliance, making exit navigable rather than terminal. We further recommend displaying visible alliance culture indicators before joining, such as conflict frequency, member retention rate, and governance style, so players can assess community fit before committing. Both features reduce exit costs without eliminating exit as a meaningful signal of community health. Together they address the screening mechanism's long-term cost: a community that can refresh its membership without catastrophic social loss is more resilient and more sustainable than one where every exit is irreversible.

\subsubsection{Formally Recognize Multiple Definitions of Success and Make Them Portable}

In SLG game communities, formal recognition systems currently track only combat power and spending-derived stats. The R1 to R5 alliance hierarchy is combat-power-adjacent and does not capture the organizational and social labor that alliances depend on to function. Management coordination, diplomatic work, event organization, and community building are invisible to the game's interface and unportable across alliances, meaning players who invest in these activities lose all recognition when they migrate to a new community.

Our data consistently showed that players pursued success through fundamentally different routes. Some optimized for strategic positioning within the spending hierarchy. Others built status through alliance management and organizational coordination. Others converted social activity and event participation into material benefits and community recognition. Others found full value in collective participation regardless of competitive outcome. These were not marginal or compensatory behaviors. They were genuine achievement pathways that alliances depended on, consistent with research on social capital and leadership in digital communities \cite{ang2010social,lisk2012leadership}. Yet none of them were tracked or recognized by the game's formal systems. Xie et al. \cite{xie2022roleseer} show that informal social roles in MMORPGs are fluid and valuable but structurally unrecognized by formal systems, and our findings extend this into an economic inequality context: the invisibility of social labor in Whiteout Survival means that non-spending players who contribute organizationally have no portable credential to show for it.  Besides, drawing on research on sense-making and role construction in online environments \cite{schiller2018inside, o2015sense}, we suggest that introducing formalized non-combat roles, such as diplomat, strategist, or logistics coordinator, could diversify status hierarchies and redistribute recognition. By visualizing and quantifying social labor, design can expand the criteria of legitimacy beyond economic or military superiority. Such mechanisms would allow players to reconfigure identity within existing hierarchies rather than being locked into marginal positions, echoing findings on identity negotiation and role flexibility in online communities \cite{guegan2015online}. 

Online gaming communities develop recognizable hierarchical structures through skill, resource control, and social influence. However, these hierarchies are often numerically quantified, reinforcing combat-based or resource-based dominance. Our interview reveal alternative status pathways consistent with research on social capital and leadership in digital communities \cite{ang2010social,lisk2012leadership}.

We therefore recommend introducing a portable Contribution Score that tracks quantifiable social behaviors, including event participation rate, new player onboarding actions, coordination activity, and attendance at collective events. This score should appear as a secondary ranking on player profiles and alliance leader-boards, and serve as an alternative criterion for role promotions alongside combat power. Critically, it should travel with the player across alliance switches and server migrations, converting informal social capital into a persistent credential rather than an alliance-specific norm that disappears on exit. This directly reduces the migration cost identified in 5.6.2, because players who have built status through social labor can carry that recognition with them rather than starting from zero in every new community. The goal is not to eliminate the spending hierarchy but to give it a parallel track that the game's own social structure already depends on and rewards informally.

Taken together, these three design directions do not aim to eliminate inequality in persistent SLG games. They aim to make it governable. Surfacing cost and interdependence information transforms opaque hierarchies into legible ones, giving players the information they need to make rational decisions about where they stand and why. Supporting low-cost exit and informed re-entry transforms a calcifying social structure into one that can renew itself without catastrophic loss. Formally recognizing multiple definitions of success transforms a single-axis spending hierarchy into a system that rewards the organizational and social labor it already depends on. Inequality in digital environments is not inherently destabilizing. Its consequences depend on whether players can understand it, navigate it, and find a meaningful place within it on their own terms.

\subsection{Limitation}

\subsubsection{Lack of Controlled Conditions}
This study is based on qualitative data collected from a naturally occurring pay-to-win game ecology. Therefore, our findings should not be interpreted as evidence of causal effects. Without controlled conditions, we cannot determine whether players’ emotional responses and behavioral adaptations were caused by inequality alone, or by the combined influence of alliance culture, server norms, interpersonal conflicts, and prior game-play experience. Our contribution is instead to show how inequality is perceived and interpreted within the complex social and economic ecology of live-service game-play.

\subsubsection{Survivor Bias and Uneven Representation}
Because players in management roles or socially central alliance positions were more likely to participate and provide detailed accounts, our dataset contains richer descriptions from socially embedded players. In contrast, highly marginalized players, early quitters, and top high-spending players are less represented. Therefore, the research results may more reflect the meaning construction process of socially active participants rather than those who have left early or those whose economic advantages prevent them from being affected by inequality pressure. 

\subsubsection{Comparison of Realistic and Virtual Data}
Although some participants reflected on the similarities or differences between their reactions in the game and the unequal conditions they experienced in reality, no systematic data on the systemic social economic status in the real world was collected. Therefore, the conclusions regarding cross-context consistency are still at the exploratory stage. Our data do not allow us to determine whether players’ in-game responses reflect stable personal orientations or context-specific strategies.

\subsection{Future Work}

Based on these limitations and our findings, we identify several directions for future work.

First, Transparency is not naturally beneficial. Future research needs to critically evaluate the role of “transparency” as a design intervention. Although clearer information about spending, rankings, and reward distribution may help players make more informed decisions, it may also intensify upward comparison or further legitimize the dominance of high-spending players. Therefore, future research needs to distinguish which types of transparency can genuinely reduce harm, and which types of transparency merely make inequality more visible without making players more capable of questioning or changing it.

Second, Our interviews suggest that players’ narratives about inequality may involve self-presentation and identity negotiation. For example, some players describe themselves as social players and say that they do not care much about combat power gaps, while still checking rankings or comparing alliance strength. Future research should not simply treat this phenomenon as inconsistency, but could further analyze how players construct their identities as “casual,” “social,” or “rational” players while continuing to participate in competitive and hierarchical systems.

Third, Future research could further examine how players exit. Since this study mainly focuses on players who remained in the game, players who left in the first few weeks or months may have experienced inequality in different ways; their absence also limits our ability to explain the relationship between inequality and churn. Future longitudinal research could track players who exit and compare the pathways of retained and churned players, in order to analyze why players leave or stay, whether and to what extent this is related to inequality, emotional tolerance thresholds, and how the balance of mobility is formed in unequal game ecosystems.

Finally, future research could further examine the relationship between “visible inequality” and “hidden structural unfairness” in game and real-world contexts. This paper mainly focuses on inequalities that players can perceive and articulate, such as leader-board gaps, power gaps, alliance resource distribution, and differences in social position. However, whether in games or in real-world social systems, unfairness does not always appear as clearly visible gaps. In games, it may be embedded in less transparent mechanisms such as algorithmic recommendation or paid progression path design. In real-world contexts, structural unfairness may also exist in implicit mechanisms such as institutional rules, information availability, and social networks. Therefore, future research could combine player interviews with game mechanics and data analysis. Such research could help HCI and game studies move beyond the level of subjective perception and further understand how inequality systems are jointly maintained through visible differences and less noticeable structural mechanisms.

\section{Conclusion}\label{sec:Conclusion}
In this study, we examined how players perceive and respond to structural inequality in the online simulation game Whiteout Survival using semi-structured interviews and in-game behavioral observations. In virtual environments does not simply produce negative experiences, rather, its meaning is actively interpreted through players’ relative positions within community hierarchies. The most salient pattern is “The Dragon Slayer Becomes the Dragon”- players who initially criticized unequal systems often came to accept, and even defend, the same structures after gaining status, resources, or social capital. This position-dependent shift demonstrates that perceived relative status plays a more obvious role than objective numerical disparity in shaping legitimacy judgments. Moreover, the transparency of game rules, structural interdependence among players, and mobility within hierarchies significantly moderate how inequality is experienced. When mechanisms are explainable, inequality becomes more psychologically tolerable and socially negotiable.

This paper challenges a one-sided view of freemium and pay-to-win games by showing that players are not simply passive victims of monetized inequality. Instead, they actively interpret, legitimize, resist, adapt to, and sometimes reproduce inequality within a persistent, alliance-based game community.

\bibliographystyle{ACM-Reference-Format}
\bibliography{references}

\newpage
\appendix

 \label{sec:Appendix}
\section{Appendix: \label{appendix}}
\appendix 

\subsection{Semi-Structured Interview Protocol} \label{app:Questionnaire}

\subsubsection{Phase 1: Perception and Attribution of Inequality}
\textit{Goal: To identify the "triggers" of awareness and how players internalize the causes of disparity.}

\begin{enumerate}
    \item Emergence of Awareness: Can you recall the first time you realized that certain players possessed "structural advantages" (advantages built into the mechanics, rather than accidental gains)?
\textit{Probing:} Please describe the specific scenario or event that triggered this realization.

    \item Causal Attribution: In your view, what are the primary drivers of these disparities? Why do certain players reach the top of the server hierarchy?
\textit{Probing:} To what extent do you attribute this to the game rules, long-term resource accumulation, alliance power structures, or platform-level monetization?

\end{enumerate}

\subsubsection{Phase 2: Dimensions of Inequality (Resources, Access, and Status)}
\textit{Goal: To explore how inequality manifests in material and social forms.}

\begin{enumerate}
    \item Resource Disparity: How do you react when you see others possessing rare resources or high-tier heroes that are practically inaccessible to you?
\textit{Probing:} Does this disparity make you feel that "effort leads to growth," or that "the gap is unbridgeable regardless of play-style"? How does this affect your judgment of the game's overall "fairness"?

    \item Social Comparison: Who do you most frequently compare yourself with—players slightly stronger than you, or the top-ranked "whales" on the server? Why?
    \item Access and Agency: Have you ever felt forced to obey others or felt a sense of dependency due to your limited power (level/combat strength)?
\textit{Probing:} Does this state of dependency feel like receiving "support" or being "controlled/marginalized"?

    \item Social Status: What specific traits determine a player’s status in your server? Does your behavior or social strategy change when you occupy different positions in the hierarchy?
    \item Institutional Fairness: How do you perceive "fairness mechanics"? Do these provide genuine equity, or do they merely mask the underlying inequality?
\end{enumerate}

\subsubsection{Phase 3: Impact on Decision-Making and Risk Behavior}
\textit{Goal: To examine how unfavorable positions alter strategic choices.}
\begin{enumerate}
    \item Risk Propensity: How do you adjust your decision-making when you realize you are at a disadvantage? Do you become more conservative to protect resources, or more adventurous ("gambling") to close the gap?
    \item Countervailing Actions: Have you observed or participated in player-led efforts to "balance the scales" (e.g., alliance treaties, tactical innovations)? Do these effectively reduce inequality?
    \item Malicious Glee (Schadenfreude): How do you feel when top-tier players who were previously aggressive or dominant suffer setbacks (e.g., losing in SvS battles or being banned)?
\textit{Probing:} Is there a sense of "poetic justice"? Do you and your allies celebrate such events?

\end{enumerate}

\subsubsection{Phase 4: Emotional Trajectories and Psychological Adaptation}

\textit{Goal: To capture the affective toll of systemic inequality.}

Emotional Impact: How does the persistent inequality shape your emotional experience of the game?

\textit{Probing:} Does it trigger anxiety and a drive to compete, or lead to numbness and "passive disengagement"?

\subsubsection{Phase 5: Virtual vs. Real-World Reflection}

\textit{Goal: To compare the player's perception of inequality across contexts.}

Contextual Mirroring: How does the inequality in this game compare to inequality in real life? Does the "gaming" context make these disparities easier or harder to accept?

\subsubsection{Phase 6: Conclusion and Design Interventions}

\textit{Goal: To solicit participant-led solutions and final thoughts.}

\begin{enumerate}
    \item Design Reform:\textbf{ }If you could modify one core mechanic to optimize the experience of inequality, what would it be?
    \item Responsibility: Who do you believe should be held responsible for the consequences of these inequalities (the players, the alliance leaders, or the game designers)?
\end{enumerate}
\end{document}